\documentstyle[11pt]{article}

\makeatletter
\@addtoreset{equation}{section}
\makeatother
\renewcommand{\theequation}{\thesection.\arabic{equation}}

\def\dalemb#1#2{{\vbox{\hrule height .#2pt
        \hbox{\vrule width.#2pt height#1pt \kern#1pt
                \vrule width.#2pt}
        \hrule height.#2pt}}}

\let\a=\alpha \let\b=\beta   \let\e=\epsilon

 \def\bd{\begin{document}} \def\ed{\end{document}}
\def\ds{\documentstyle} \let\fr=\frac \let\bl=\bigl \let\br=\bigr
\let\Br=\Bigr \let\Bl=\Bigl 
\let\bm=\bibitem
\let\na=\nabla
\let\pa=\partial \let\ov=\overline
\let\ul=\underline 
\newcommand{\be}{\begin{equation}} 
\newcommand{\ee}{\end{equation}} 
\def\ve{\varepsilon}
\def\ba{\begin{array}}
\def\ea{\end{array}}
\def\ft#1#2{{\textstyle{{\scriptstyle #1}\over {\scriptstyle #2}}}}
\def\fft#1#2{{#1 \over #2}}
\def\del{\partial}
\def\sst#1{{\scriptscriptstyle #1}}
\def\oneone{\rlap 1\mkern4mu{\rm l}}
\def\e7{E_{7(+7)}}
\def\td{\tilde}
\def\bog{Bogomol'nyi\ }
\def\ss{Scherk-Schwarz\ }
\def\kk{Kaluza-Klein\ }
\def\ads{anti-de Sitter\ }
\def\Z{\rlap{\ssr Z}\mkern3mu\hbox{\ssr Z}}
\def\R{\rlap{\rm I}\mkern3mu{\rm R}}
\newcommand{\ho}[1]{$\, ^{#1}$}
\newcommand{\hoch}[1]{$\, ^{#1}$}
\newcommand{\bea}{\begin{eqnarray}} 
\newcommand{\eea}{\end{eqnarray}} 
\newcommand{\ra}{\rightarrow}
\newcommand{\lra}{\longrightarrow}
\newcommand{\Lra}{\Leftrightarrow}
\newcommand{\ap}{\alpha^\prime}
\newcommand{\bp}{\tilde \beta^\prime}
\newcommand{\tr}{{\rm tr} }
\newcommand{\Tr}{{\rm Tr} } 
\newcommand{\NP}{Nucl. Phys. }
\newcommand{\tamphys}{\it $^{(2)}$ Center for Theoretical Physics,
Texas A\&M University, College Station, Texas 77843}

\newcommand{\auth}{P.M. Cowdall$^{(1)}$, H. L\"u$^{(2,3)}$, 
C.N. Pope$^{(2,3,\dagger)}$, K.S.
Stelle$^{(4,\star)}$ and P.K. Townsend$^{(1)}$
}

\begin{document}
\begin{flushright}
\hfill{CTP TAMU-26/96}\\
\hfill{DAMTP-R/96/32}\\
\hfill{Imperial/TP/95-96/68}\\
\hfill{SISSA 124/96/EP}\\
\hfill{hep-th/9608173}\\
\end{flushright}

\vspace{5pt}

\begin{center}
{\large {\bf Domain Walls in Massive Supergravities}}

\vspace{15pt}

\auth

\vspace{10pt}

{\it $^{(1)}$ DAMTP, University of Cambridge,
Silver St., Cambridge CB3 9EW, U.K.}

\vspace{6pt}

{\tamphys}

\vspace{6pt}

{\it $^{(3)}$ SISSA, Via Beirut No. 2-4, 34013 Trieste, Italy }

\vspace{6pt}

{\it $^{(4)}$ The Blackett Laboratory, Imperial College, Prince Consort Road,
London SW7 2BZ}

\vspace{15pt}

\underline{ABSTRACT}
\end{center}

     We show how toroidally-compactified eleven-dimensional supergravity can
be consistently truncated to yield a variety of maximally-supersymmetric
{\it massive} supergravities in spacetime dimensions $D\le 8$.  The mass
terms arise as a consequence of making a more general ansatz than that in
usual Kaluza-Klein dimensional reduction, in which one or more axions are
given an additional linear dependence on one of the compactification
coordinates.  The lower-dimensional theories are nevertheless consistent
truncations of eleven-dimensional supergravity. Owing to the fact that the
generalised reduction commutes neither with U-duality nor with ordinary
dimensional reduction, many different massive theories can result.  The
simplest examples arise when just a single axion has the additional linear
coordinate dependence.  We find five inequivalent such theories in $D=7$,
and 71 inequivalent ones in $D=4$.  The massive theories admit no
maximally-symmetric vacuum solution, but they do admit $(D-2)$-brane
solutions, {\it i.e.}\ domain walls, which preserve half the supersymmetry. 
We present examples of these solutions, and their oxidations to $D=11$. 
Some of the latter are new solutions of $D=11$ supergravity.

{\vfill\leftline{}\vfill
\vskip	10pt
\footnoterule {\footnotesize
\hoch{\dagger}	Research supported in part by DOE  Grant
DE-FG05-91-ER40633 and \vskip	-12pt}  \vskip	8pt {\footnotesize  EC Human
Capital and Mobility Programme under contract ERBCHBGCT920176. 
\vskip	-12pt}  \vskip	10pt {\footnotesize 
\hoch{\star} Research supported in part by the Commission of the  European
Communities under contract \vskip	-12pt}  \vskip	8pt {\footnotesize
SCI*-CT92-0789.} } 

\pagebreak
\setcounter{page}{1}     

\section{Introduction}

     M-theory or string theory, and their dimensional reductions, admit a
multitude of $p$-brane soliton solutions.  These can be understood from a
variety of perspectives.  In particular, one way to classify them is by
giving their explicit construction in the various lower-dimensional
supergravities obtained by dimensional reduction from M-theory or string
theory.  An isotropic $p$-brane solution in $D$ dimensions is supported by
either an electric or a magnetic charge carried by an antisymmetric-tensor
field strength in the theory; a rank $p+2$ tensor in the electric case, or a
rank $(D-p-2)$ tensor in the magnetic case.  The numbers of field strengths
with ranks $\le 3$ grow rapidly with decreasing $D$, and, correspondingly,
the numbers of $p$-brane solutions for each $p$ increase also.  These
multiplicities of solutions are classified by the U duality symmetries of
the corresponding supergravity theories, (or more precisely, by the Weyl
groups of the U duality groups \cite{lpsweyl}).  Another approach to the
classification of $p$-brane solitons is to follow their evolution through
the various steps of dimensional reduction \cite{ht}, from a set of
higher-dimensional solutions of the original M-theory or string theory. 
Some, but not all, of these higher-dimensional solutions are themselves
$p$-brane solitons. 

     An essential ingredient that allows one to give these two kinds of
description for the $p$-brane spectrum is the fact that the Kaluza-Klein
reduction of the higher-dimensional theory is a {\it consistent truncation}.
 In particular, this means that all solutions of the lower-dimensional
theory are necessarily also solutions of the higher-dimensional one.  A
further feature is that (down to $p=D-3$, as we shall see), from a given
starting-point in the higher dimension, there is a {\it unique theory}
relevant to the solutions in the lower dimension.  This may, without loss of
generality, be taken to be the maximal supergravity in that dimension, as
all lesser supergravities may be taken to be consistent truncations of a
maximal theory. This unique theory has a high degree of symmetry, namely the
U duality referred to above, which incorporates the multiplicity of
reduction routes one may use to arrive at a given pair $(D,d)$ of spacetime
and worldvolume dimensions. 

     These different routes employ two basic types of reduction of $p$-brane
solutions.  The first of the two reduction procedures makes use of the
Poincar\'e symmetry of such solutions on the $p$-brane worldvolumes, and
results in a simultaneous reduction of the spacetime and worldvolume
dimensions, from $(D,d)$ to $(D-1,d-1)$ \cite{dhis,dl,lpss1}.  The second
reduction procedure involves first constructing a ``stack'' of parallel
$p$-branes.  By taking the continuum limit as this stack becomes densely
packed, a translational invariance is generated in a direction transverse to
the worldvolume of the $p$-brane, and this can then be used for the
Kaluza-Klein reduction \cite{lpsvert,ht}. In this case, $(D,d)$ is reduced
to $(D-1,d)$. The two reduction procedures have been called ``diagonal'' and
``vertical,'' respectively.  In performing a vertical reduction, taking the
continuum limit of a stack of $p$-branes amounts to replacing a sum by an
integral within a multi-center solution of Laplace's equation in the
transverse space, effectively reducing by one the dimension of the space in
which the Laplacian acts. Thus, under vertical reduction, $(D,d)$ is reduced
to $(D-1,d)$. 

     This process of generating harmonic functions in lower transverse
dimensions continues uneventfully until one makes the reduction from the
harmonic function $1/r$ in three dimensions to $\log r$ in two dimensions,
corresponding to the reduction from a $(D,D-3)$ solution to a $(D-1,D-3)$
solution.\footnote{We shall refer generally to solutions of this latter
sort, where the worldvolume dimension $d=p+1$ is 2 less than the spacetime
dimension as ``$(D-3)$-branes,'' regardless of whether one is in $D$
spacetime dimensions or not.} This step has two related features.  The first
is that the integral $\int_{-L}^L dz (r^2+z^2)^{-1/2}$ producing the $\log
r$ harmonic function gives rise to an additive $2\log L$ divergence as
$L\rightarrow\infty$ that needs to be renormalised away.  The second feature
is that the $\log r$ harmonic function no longer dies off at transverse
infinity.  Its derivative, which governs the form of the field strength
supporting the solution, still does fall off at transverse infinity,
however. 

     Continuing on through one more step, the reduction from a
two-dimensional to a one-dimensional transverse space involves the integral
$\int_{-L}^L dz \log(y^2 + z^2)$ which again produces an additive
divergence, this time of the form $4L \log L -4L$.  After renormalising this
away, the resulting harmonic function $H(y)$ is linear in the one remaining
transverse-space coordinate $y$.  Not only does this not vanish at
transverse infinity, but its derivative is now a non-vanishing constant as
well.  This has the consequence that the corresponding 1-form field strength
supporting the continuous stack of $(D-3)$-branes has the form (still in the
two transverse dimensions) $F_m= -\epsilon_{mn} \del_n H(y)$, and thus
$F_z$, its component in the compactification direction, is a constant.  This
is a new phenomenon in the process of vertical reduction, which we shall
study in detail in the rest of this paper.  In particular, this has the
consequence that the 0-form potential, or ``axion,'' for the field strength
$F_m$ must itself be linearly dependent on $z$.  (The occurrence of membrane
solutions in $D=4$ resulting from vertical reduction of a 5-brane of the
heterotic string wrapped around a 3-torus was observed in \cite{detal},
where it was also noted that the usual Kaluza-Klein ansatz is not
sufficient, since it does not give rise to the necessary cosmological term
in the reduced $D=4$ theory.) 

     Normally, in Kaluza-Klein dimensional reduction, all of the
higher-dimensional fields are taken to be independent of the
compactification coordinate $z$.  Indeed, consistency of the truncation is
assured by retaining all fields subject to this restriction.  At first
sight, one might think that a consistent truncation would no longer be
possible if any dependence on the compactification coordinate were retained.
 However, the supergravity theories whose dimensional reduction we are
considering have a special feature that the axionic field that is given the
above $z$ dependence always appears in the Lagrangian covered by a
derivative. Consequently, after insertion of the ansatz, the
higher-dimensional Lagrangian and also the equations of motion will still be
independent of $z$, thus permitting a consistent truncation even when we
allow a more general Kaluza-Klein ansatz in which such an axion has a linear
$z$ dependence.  In fact, what vertical reduction has produced in this case
is an instance of Scherk-Schwarz type Kaluza-Klein reduction
\cite{ss,bdgpt}.\footnote{This instance of the Scherk-Schwarz procedure uses
the shift symmetry of the selected axionic field as the global symmetry to
be supplied with a $z$-dependent gauge parameter in the Kaluza-Klein
reduction.}  However, unlike the original Scherk-Schwarz procedure, which
was designed to give a mass to the gravitino, thus breaking supersymmetry,
it turns out in the present case that supersymmetry remains unbroken, but
that certain bosonic fields become massive instead.  The set of massive
fields always includes the Kaluza-Klein vector potential arising from the
reduction of the metric, the additional degree of freedom being supplied by
the dimensionally-reduced axion that was given the linear $z$ dependence in
the higher dimension. At the same time, a cosmological-type term is
generated in the lower-dimensional theory, having the general form of
$\sqrt{-g}$ times an exponential of dilatonic scalar fields.  The first
example of this kind of Scherk-Schwarz procedure arose in the dimensional
reduction of the type IIB theory \cite{bdgpt}, where it was shown that the
resulting nine-dimensional theory is related by $T$ duality to the ordinary
Kaluza-Klein reduction of massive IIA supergravity \cite{r}. 
     
     In the rest of this paper, we shall study the above process of
Scherk-Schwarz dimensional reduction in supergravity theories, and the
consequences of this for $p$-brane solutions.  In doing so, our principal
focus will be on the consistent reductions of the theories themselves,
rather than on the relations between solutions of the reduced theories and
solutions in higher dimensions.  A particularly compelling reason for
placing the emphasis on theories rather than on solutions is that, as we
shall show, the process of Scherk-Schwarz dimensional reduction of a given
theory can be done in as many ways as there are axions in the
higher-dimensional theory.\footnote{We are using the term ``axion'' here to
denote any scalar field that has a shift symmetry under which all other
fields are inert, and thus can, after appropriate field redefinitions and
integrations by parts, be covered everywhere by a derivative.  The axion
fields in the supergravity theories that we are considering all arise as the
0-form potential endpoints of dimensional reductions of higher-rank
potentials (except for the axion in type IIB supergravity).} This situation
is to be contrasted with that prevailing for $p$-branes with $p\le(D-3)$,
where there is always an essentially {\it unique} reduced supergravity
theory ({\it i.e.}\  unique up to standard duality transformations). The
various pathways of dimensional reduction above this $p=(D-3)$ ``barrier''
thus give rise to a multiplicity of different solutions to a single theory,
with the multiplicity being classified by the Weyl group of the U-duality
symmetry. Penetrating this $p=(D-3)$ barrier by the various forms of
Scherk-Schwarz reduction that we shall discuss gives rise, not to a
multiplicity of solutions within a given theory, but to a multiplicity of
different reduced theories (with only a trivial $Z_2$ residual classifying
symmetry for the resulting $(D-2)$-brane solutions so obtained). 

     In the next section, we shall describe in detail the process of
Scherk-Schwarz dimensional reduction, showing that it commutes neither with
ordinary Kaluza-Klein reduction, nor with the U-duality symmetries. This
latter point has the consequence that every axionic scalar gives rise to its
own distinct Scherk-Schwarz reduced theory. Furthermore, the
non-commutativity with ordinary Kaluza-Klein reduction means that the total
number of distinct massive supergravities obtained by these procedures is
the cumulative total of those already obtained in one higher dimension, plus
the further examples obtained by Scherk-Schwarz reduction in the final step.
 We shall discuss in detail the massive theories in the specific cases of
$D=8$ and $D=7$, showing that there are two such distinct massive theories
in $D=8$, and six distinct theories in $D=7$. 

\section{Scherk-Schwarz dimensional reduction}

     The highest-dimensional example of a massive supergravity theory is the
massive IIA theory in $D=10$, which was constructed in \cite{r}.  It was
shown in \cite{bdgpt} that its Kaluza-Klein reduction to $D=9$ coincides, up
to a $T$-duality transformation, with the massive theory obtained by
Scherk-Schwarz reduction of the type IIB theory.  Subsequent steps of
Kaluza-Klein reduction will give rise to a single massive supergravity in
each lower dimension.  None of these examples can be obtained by any
combination of \ss and \kk reductions of $D=11$ supergravity.  All other
examples of massive supergravities of the kind that we are considering in
this paper can, however, be obtained from $D=11$.  Accordingly, we shall
present just the cases derived from the $D=11$ theory in detail,
concentrating principally on the bosonic sectors of the theories. 

    It is convenient to parametrise the standard \kk dimensional reductions
of $D=11$ supergravity using the formalism presented in \cite{lpsol}, which
arrives at $D$ dimensions by means of a succession of 1-step reductions,
rather than by a direct descent from 11 to $D$ dimensions.  The usual \kk
reduction procedure involves taking the higher-dimensional fields to be
independent of the compactification coordinate $z$.  A convenient
parametrisation for the metric and for an $n$-rank potential $A_n$, is then
as follows: 
\bea
ds^2(x,z) &\longrightarrow & e^{2\a\varphi(x)} ds^2(x) + e^{-2(\sst D -2) \a
\varphi(x)} (dz+ {\cal A}_{\sst M}(x) dx^{\sst M})^2 \ ,\nonumber\\
A_n(x,z) &\longrightarrow & A_n(x) + A_{n-1}(x) \wedge dz \ ,\label{kkans}
\eea
where the fields in $(D+1)$ dimensions on the left are reduced to fields in
$D$ dimensions on the right.  The constant $\a$ is given by $\a=
(2(D-1)(D-2))^{-1/2}$.  This parametrisation of the metric is chosen so that
the Einstein action $\sqrt{-g} R$ in the higher dimension reduces to
$\sqrt{-g} R$ in the lower dimension.  The field strengths are best
expressed in terms of their orthonormal-frame components, obtained by
contracting with vielbeins, since these most easily allow the forms of their
lower-dimensional kinetic terms to be read off. Thus we find that a field
strength $F_{n+1}(x,z)=dA_n(x,z)$ in the higher dimension reduces to 
\be
dA_n(x) - dA_{n-1}(x)\wedge {\cal A}(x) + dA_{n-1}(x)
\wedge (dz + {\cal A})\equiv F_{n+1}(x) + F_n(x)\wedge (dz+{\cal A})\ ,
\ee
showing how Chern-Simons modifications of the form $F_{n+1} = dA_n -
dA_{n-1} \wedge {\cal A}$ are developed as the reduction proceeds. The
resulting bosonic Lagrangian for maximal supergravity in $D$ dimensions was
obtained using this formalism in \cite{lpsol}, and is presented in the
Appendix. 

In the context of $D=11$ supergravity reductions that we are considering,
the highest dimension in which there is an axion available for performing a
\ss reduction is $D=9$. We shall begin by considering this example in some
detail. 

\subsection{\ss reduction of $D=9$ supergravity}

{}From the formulae collected in the
Appendix, the Lagrangian for massless $D=9$ maximal supergravity is given by
\bea {\cal L} &=& eR -\ft12e (\del\phi_1)^2 -\ft12e (\del\phi_2)^2 -
\ft12 e e^{\phi_1+\ft3{\sqrt7}\phi_2} (\del\chi)^2 -\ft1{48} e
e^{\vec a\cdot \vec\phi} (F_4)^2\nonumber\\
&&- \ft12 e e^{\vec a_1\cdot \vec\phi} (F_3^{(1)})^2 -\ft12 e
e^{\vec a_2\cdot \vec\phi} (F_3^{(2)})^2 
-\ft14 e e^{\vec a_{12}\cdot \vec\phi} (F_2^{(12)})^2 -\ft14 e
e^{\vec b_1\cdot\vec\phi} ({\cal F}_2^{(1)})^2 \label{d9lag}\\ 
&&- \ft14 e e^{\vec b_2\cdot \vec\phi} ({\cal F}_2^{(2)})^2  
-\ft12 \td F_4\wedge \td F_4 \wedge A_1^{(12)} -
\td F_3^{(1)} \wedge \td F_3^{(2)} \wedge A_3\ ,\nonumber
\eea
where we have defined $\vec\phi=(\phi_1,\phi_2)$, and $\chi={\cal 
A}_0^{(12)}$.  We have made the dilaton coupling within the scalar-manifold
sector explicit, while in the rest of the Lagrangian the dilaton vectors are
as given in the Appendix. The Chern-Simons modifications to the field
strengths are given by 
\bea 
F_4&=&\td F_4 - \td F_3^{(1)}\wedge {\cal A}_1^{(1)} -
\td F_3^{(2)}\wedge {\cal A}_1^{(2)} +\chi \td F_3^{(1)}
\wedge {\cal A}_1^{(2)} - \td F_2^{(12)}
\wedge {\cal A}_{1}^{(1)}
\wedge {\cal A}_1^{(2)}\ ,\nonumber\\ 
F^{(1)}_3 &=& \td F^{(1)}_3 - \td
F_2^{(12)} \wedge {\cal A}_1^{(2)}\ ,
\nonumber\\ 
F_3^{(2)} &=& \td F_3^{(2)} + F_2^{(12)}\wedge {\cal
A}_1^{(1)} -
\chi \td F^{(1)}_3 
\ ,\label{csterms}\\ 
F_2^{(12)} &=& \td F^{(12)}_2\ ,\quad {\cal
F}_2^{(1)} = {\cal F}_2^{(1)} -d\chi\wedge{\cal A}_1^{(2)}\ ,\quad {\cal
F}_2^{(2)} = \td {\cal F}_2^{(2)}\ , \quad {\cal F}_1^{(12)} = 
d\chi\ .\nonumber
\eea

     As it stands, we can see from (\ref{csterms}) that  the axion $\chi$
appears without yet being covered by a derivative in certain  places in the
Lagrangian.  However, it is easy to remedy this by performing the following
field redefinition: 
\be
A_2^{(2)}\longrightarrow A_2^{(2)} + \chi A_2^{(1)} \ ,\label{9redef}
\ee
after which the Chern-Simons modifications (\ref{csterms}) become
\bea
F_4&=&\td F_4 - \td F_3^{(1)}\wedge {\cal A}_1^{(1)} -
\td F_3^{(2)}\wedge {\cal A}_1^{(2)} -d\chi\wedge A_2^{(1)}
\wedge {\cal A}_1^{(2)} - \td F_2^{(12)}
\wedge {\cal A}_{1}^{(1)}
\wedge {\cal A}_1^{(2)}\ ,\nonumber\\
F^{(1)}_3 &=& \td F^{(1)}_3 - \td
F_2^{(12)} \wedge {\cal A}_1^{(2)}\ ,
\nonumber\\
F_3^{(2)} &=& \td F_3^{(2)} + F_2^{(12)}\wedge {\cal
A}_1^{(1)} +
d\chi \wedge A_2^{(1)}
\ ,\label{newcsterms}\\
F_2^{(12)} &=& \td F^{(12)}_2\ ,\quad {\cal
F}_2^{(1)} = {\cal F}_2^{(1)} -d\chi\wedge{\cal A}_1^{(2)}\ ,\quad {\cal
F}_2^{(2)} = \td {\cal F}_2^{(2)}\ , \quad {\cal F}_1^{(12)} = 
d\chi \ .\nonumber
\eea
One may argue on general grounds that, having established that the axion
$\chi$ can be covered by a derivative everywhere in the bosonic sector, the
same will also be true in the fermionic sector.  The reason for this is that
the full set of scalar fields in maximal supergravity ({\it i.e.}\ both
axions and dilatons) are described by a coset manifold, and that their
couplings to the fermions occur {\it via} the vielbein on this coset
manifold.  We have seen above that the metric on the coset manifold can be
cast into a form where there is no dependence on the axion $\chi$, and
therefore it follows that a vielbein basis with this same property can also
be chosen.  Then, the entire supergravity Lagrangian will involve $\chi$
covered everywhere by a derivative. 

     We are now in a position to proceed with the \ss dimensional reduction
to $D=8$.  Denoting the compactification coordinate by $z$, we make the
usual \kk ansatz (\ref{kkans}) that all the lower-dimensional fields are
independent of $z$, with the exception of the axion $\chi$, for which the
ansatz will be 
\be
\chi(x,z)\longrightarrow m z + \chi(x)\ .\label{ssans}
\ee
Note that the vielbein components of the field strength for $\chi$ can then
be read off from $d\chi \rightarrow d\chi - m {\cal A} + m(dz+{\cal A})$,
implying in particular that there will be a ``0-form field strength'' $m$,
which will give rise to a cosmological-type term in the lower dimension. It
is manifest that the higher-dimensional equations of motion, after
substitution of the ansatz, will be independent of $z$, and hence it follows
that this truncation to $D=8$ will be consistent, since all the
$z$-independent degrees of freedom are being retained.  Furthermore, it
follows for the same reason that the $D=8$ equations of motion are derivable
from the $D=8$ Lagrangian that one obtains by simply substituting the ansatz
into the $D=9$ Lagrangian. 

     Performing the substitution of the ansatz into (\ref{d9lag}), we obtain
the following $D=8$ Lagrangian: 
\bea
e^{-1}\cal L &=&  R -\ft12  (\del\phi_1)^2 -\ft12  (\del\phi_2)^2 
-\ft12  (\del\phi_3)^2 -
\ft12 e^{\vec b_{12}\cdot \vec \phi}(\del \chi - m 
{\cal A}_1^{(3)})^2\nonumber\\
&&-\ft12 e^{\vec b_{13}\cdot \vec \phi}
(\del {\cal A}_0^{(13)} + m {\cal A}_1^{(2)})^2
-\ft12
e^{\vec b_{23}\cdot \vec \phi}(\del {\cal A}_0^{(23)})^2 -\ft12
e^{\vec a_{123}\cdot \vec \phi}(\del A_0^{(123)})^2
\nonumber\\
&&-\ft1{48} e^{\vec a\cdot\vec \phi} (F_4- m A_2^{(1)} \wedge {\cal
A}_1^{(2)}\wedge {\cal A}_1^{(3)})^2 
-\ft1{12} e^{\vec a_1\cdot
\vec\phi}(F_3^{(1)})^2 \nonumber\\
&&-\ft1{12} e^{\vec a_2\cdot
\vec\phi}(F_3^{(2)}+ m A_2^{(1)}\wedge {\cal A}_1^{(3)})^2 
-\ft1{12} e^{\vec a_3\cdot
\vec\phi}(F_3^{(3)} + m A_2^{(1)} \wedge {\cal A}_1^{(2)}  )^2\nonumber\\ 
&&-\ft14 e^{\vec a_{12}\cdot \vec \phi}(F_2^{(12)})^2 
-\ft14 e^{\vec a_{13}\cdot \vec \phi}(F_2^{(13)})^2
-\ft14 e^{\vec a_{23}\cdot \vec \phi}(F_2^{(23)}+m A_2^{(1)})^2
\nonumber\\
&&-\ft14 e^{\vec b_{1}\cdot \vec \phi}({\cal F}_2^{(1)}
- m {\cal A}_1^{(2)} \wedge {\cal A}_1^{(3)})^2
-\ft14 e^{\vec b_{2}\cdot \vec \phi}({\cal F}_2^{(2)})^2
-\ft14 e^{\vec b_{3}\cdot \vec \phi}({\cal F}_2^{(3)})^2 \nonumber\\
&& - \ft12 m^2 e^{\vec b_{123}\cdot\vec\phi} + e^{-1}\, 
{\cal L}_{FFA}\ ,
\label{massived8}
\eea
where ${\cal L}_{FFA}$ is the same as the term given in (\ref{ffaterms}) for
massless $D=8$ supergravity, and the dilaton vectors are given by
(\ref{dilatonvec}) for the case $D=8$.  The Chern-Simons modifications for
the field strengths appearing in (\ref{massived8}) are essentially those
given by (\ref{csterms}) for the massless $D=8$ theory, after taking into
account the field redefinition (\ref{9redef}).  The penultimate term in
(\ref{massived8}) is a cosmological-type term, whose dilaton vector $\vec
b_{123}$ is given in (\ref{dilatonvec}). 

    It is now apparent that some of the fields in the $D=8$ theory that we
have obtained by \ss reduction have acquired masses.  Specifically, we see
that after performing the following gauge transformations: 
\bea
{\cal A}_1^{(3)}&\longrightarrow & {\cal A}_1^{(3)} +\fft1{m} 
d \chi\ ,\nonumber\\
{\cal A}_1^{(2)}&\longrightarrow &{\cal A}_1^{(2)} -\fft1{m}
d {\cal A}_0^{(13)}\ ,\label{gauge}\\
A_2^{(1)}&\longrightarrow & A_2^{(1)} -\fft1{m} 
d A_1^{(23)}\ ,\nonumber
\eea
the fields ${\cal A}_1^{(3)}$, ${\cal A}_1^{(2)}$ and $A_2^{(1)}$ absorb the
fields $\chi$, ${\cal A}_0^{(13)}$ and $A_1^{(23)}$ respectively. Thus
${\cal A}_1^{(3)}$, ${\cal A}_1^{(2)}$ and $A_2^{(1)}$ become massive in the
\ss reduction process. 

    This example of \ss reduction from $D=9$ to $D=8$ illustrates the
general pattern of mass generation.  Firstly, the \kk vector potential
arising in the reduction (${\cal A}_1^{(3)}$ in this example) always
acquires a mass, absorbing the dimensional reduction of the axion that was
used in the \ss reduction.  Secondly, if this axion appears in any bilinear
term in the Chern-Simons modifications for any of the higher-dimensional
fields {\it e.g.}\ $H=dB + d\chi\wedge A +\cdots $, then the corresponding
field $A$ acquires a mass also, by absorbing the field $C$ obtained from the
dimensional reduction of $B$, $B(x,z)\rightarrow B(x) + C(x)\wedge dz$.
There is also a third kind of mass generation that can arise, which is not
present in the above example. This can happen if the axion that is used in
the \ss reduction is one of the $A_0^{(ijk)}$ fields coming from the
dimensional reduction of the 3-form potential in $D=11$.  In this case, we
see from (\ref{ffaterms}) that it will give rise to a kind of topological
mass term in the lower dimension.  The first example where such an axion
exists is in $D=8$, and we shall discuss this case in more detail below. 

     It is worth emphasising that the massive theory in $D=8$ that we have
obtained above is quite distinct from the $D=8$ massive theory that is
obtained by performing a standard \kk reduction of the massive IIA theory
occurring in $D=10$ \cite{r}.  In this $D=10$ theory, the only massive field
is a 3-form field strength.  Upon performing two steps of \kk reduction,
this gives one massive 3-form, two massive 2-forms, and one massive 1-form
in $D=8$.  By contrast, the theory described by (\ref{massived8}) has only a
massive 3-form and two massive 2-form field strengths.  Thus the two
theories are manifestly inequivalent. 

     This inequivalence of the two $D=8$ massive theories illustrates
another important feature of the \ss reduction procedure, namely that it
does not commute with ordinary \kk reduction.  To see this, we note that we
could alternatively have obtained the theory (\ref{massived8}) by performing
first an ordinary \kk reduction of the type IIB theory, followed by a \ss
reduction step. On the other hand, the two-step reduction of the massive IIA
theory could alternatively have been obtained by performing first a \ss
reduction of the IIB theory \cite{bdgpt}, followed by an ordinary \kk
reduction step.  The inequivalence of the two massive theories in $D=8$ thus
demonstrates the non-commutativity of the \ss and ordinary \kk reductions. 

     It is of interest also to examine the surviving symmetries in the \ss
reduced theory.  Under ordinary \kk reduction, the $GL(2,\R)$ Cremmer-Julia
symmetry group in $D=9$ enlarges to $SL(2,\R)\times SL(3, \R)$ in $D=8$
\cite{cj}.  Since the \ss procedure gives a non-vanishing constant value to
the internal component of the 1-form field strength ${\cal F}_1^{(12)}$, we
can expect that the surviving symmetry group in $D=8$ will be smaller than
$SL(2,\R)\times SL(3, \R)$.  As in the massless theory cases, we can study
this symmetry by looking at its action on the scalar manifold describing the
dilatonic and axionic sector of the theory.  In particular, we can see from
(\ref{massived8}) that the residual symmetry transformations must leave the
cosmological term invariant, and thus the component of the dilatonic scalars
$\vec\phi$ that is parallel to $\vec b_{123}$ must be inert.  Now we find
that $\vec b_{123}\cdot \vec b_{23}=0$, while  $\vec b_{123}\cdot \vec
a_{123}\ne 0$. It therefore follows that there is an $SL(2,\R)$ invariance
involving the axion ${\cal A}_0^{(23)}$, together with the linear
combination of dilatons $\varphi\equiv -\ft12\vec b_{23}\cdot \vec\phi$,
since they are described simply by the terms $-\ft12(\del\varphi)^2 -\ft12
e^{-2\varphi} (\del {\cal A}_0^{(23)})^2$ in the scalar Lagrangian.  In
addition, there is an abelian $\R$ symmetry under which only the axion
$A_0^{(123)}$ transforms, by a constant shift $A_0^{(123)}\rightarrow
A_0^{(123)}+ {\rm const}$. Furthermore, we find that there is another $\R$
symmetry, corresponding to a constant shift of the linear combination of the
dilatons that is orthogonal both to $\vec b_{123}$ and to $\vec b_{23}$. 
Under this transformation, the various gauge potentials must scale by
appropriate exponentials of the shift parameter, so as to compensate for the
scalings of the dilaton prefactors in their kinetic terms.  A non-trivial
check on the consistency of these scalings is that the dilaton vectors
associated with the mass terms for the massive fields and also the dilaton
vectors associated with their kinetic terms must have common projections
onto the direction orthogonal to $\vec b_{123}$ and $\vec b_{23}$.  Thus, in
all, we find a symmetry 
\be
SL(2,\R)\times\R\times\R \label{d8sym}
\ee
of the massive theory described by (\ref{massived8}). 

\subsection{Scherk-Schwarz reduction of $D=8$ supergravity}

     The Lagrangian for the bosonic sector of massless $D=8$ maximal 
supergravity, in the notation given in the Appendix, is given by
\bea
e^{-1}\cal L &=&  R -\ft12  (\del\phi_1)^2 -\ft12  (\del\phi_2)^2 
-\ft12  (\del\phi_3)^2 -
\ft12 e^{\vec b_{12}\cdot \vec \phi}({\cal F}_1^{(12)})^2\nonumber\\
&&-\ft12 e^{\vec b_{13}\cdot \vec \phi}
({\cal F}_1^{(13)})^2
-\ft12
e^{\vec b_{23}\cdot \vec \phi}({\cal F}_1^{(23)})^2 -\ft12
e^{\vec a_{123}\cdot \vec \phi}(F_1^{(123)})^2
-\ft1{48} e^{\vec a\cdot\vec \phi} F_4^2 \nonumber\\
&&-\ft1{12} e^{\vec a_1\cdot
\vec\phi}(F_3^{(1)})^2 -\ft1{12} e^{\vec a_2\cdot
\vec\phi}(F_3^{(2)})^2 -\ft1{12} e^{\vec a_3\cdot
\vec\phi}(F_3^{(3)})^2 
-\ft14 e^{\vec a_{12}\cdot \vec \phi}(F_2^{(12)})^2 
\nonumber\\
&&-\ft14 e^{\vec a_{13}\cdot \vec \phi}(F_2^{(13)})^2
-\ft14 e^{\vec a_{23}\cdot \vec \phi}(F_2^{(23)})^2
-\ft14 e^{\vec b_{1}\cdot \vec \phi}({\cal F}_2^{(1)})^2\nonumber\\
&&-\ft14 e^{\vec b_{2}\cdot \vec \phi}({\cal F}_2^{(2)})^2
-\ft14 e^{\vec b_{3}\cdot \vec \phi}({\cal F}_2^{(3)})^2
+ e^{-1}\, {\cal L}_{FFA}\ .
 \label{massless8}
\eea
There are now four axions in total, namely ${\cal A}_0^{(12)}$, ${\cal
A}_0^{(13)}$, ${\cal A}_0^{(23)}$ and $A_0^{(123)}$.  The complete
expressions for the Chern-Simons modifications to the field strengths, given
by (\ref{A.6}), are quite involved, and we shall not write them  out
explicitly.  As we saw in the previous subsection, the basic structure of
the \ss reduced theory is governed by the bilinear terms in those
Chern-Simons modifications that involve the axion being used for the
reduction process.  Thus, for the purposes of determining the spectrum of
massive and massless fields in the \ss reduced theories, it suffices to
consider only the bilinear Chern-Simons modifications involving the four
axions. From the general results of \cite{lpsol}, summarised in the
Appendix, we therefore have: 
\bea
{\cal F}_1^{(12)}& =&\td{\cal F}_1^{(12)}\ ,\qquad\qquad\quad 
{\cal F}_1^{(13)} =\td{\cal F}_1^{(13)} -{\cal A}_0^{(23)} 
\td {\cal F}_1^{(12)}\ ,\qquad
{\cal F}_1^{(23)} =\td{\cal F}_1^{(23)}\ , \nonumber\\
{\cal F}_2^{(1)} &=& \td {\cal F}_2^{(1)}+ {\cal A}_1^{(2)}\wedge
\td {\cal F}_1^{(12)} +{\cal A}_1^{(3)}\wedge \td {\cal F}_1^{(13)} +
\cdots\ ,\nonumber\\
{\cal F}_2^{(2)}& =& \td {\cal F}_2^{(2)} + {\cal A}_1^{(3)} 
\wedge \td {\cal F}_1^{(23)}\ ,\nonumber\\ 
{\cal F}_2^{(3)} &=& \td {\cal F}_2^{(3)} \ ,\qquad F_1^{(123)}=
\td F_1^{(123)}\ ,\nonumber\\
F_2^{(12)} &=& \td F_2^{(12)} + {\cal A}_1^{(3)}\wedge \td F_1^{(123)}
\ ,\nonumber\\ 
F_2^{(13)}& =& \td F_2^{(13)} - {\cal A}_0^{(23)} \td F_2^{(12)} -
{\cal A}_1^{(2)} \wedge \td F_1^{(123)}\ ,
\nonumber\\
F_2^{(23)}& = &\td F_2^{(23)} - {\cal A}_0^{(12)} \td F_2^{(13)} +
{\cal A}_0^{(13)} \td F_2^{(12)}+  {\cal A}_1^{(1)} \wedge \td F_1^{(123)}
\ ,\label{bilin8}\\
F_3^{(1)}&=& \td F_3^{(1)} +\cdots\ ,\qquad 
F_3^{(2)}= \td F_3^{(2)} -{\cal A}_0^{(12)} \td F_3^{(1)}+\cdots\ ,\nonumber\\
F_3^{(3)} &=& \td F_3^{(3)}-{\cal A}_0^{(13)} \td F_3^{(1)}
-{\cal A}_0^{(23)} \td F_3^{(2)}+ \cdots\ ,\nonumber\\
F_4&=& \td F_4 +\cdots\ ,\nonumber
\eea
where the dots $\cdots$ indicate that higher-order terms, or bilinear terms
that do not involve the axions, have been omitted. 

     Let us now consider the various massive $D=7$ supergravity theories
that one can obtain by carrying out a \ss reduction step using one or
another of the four axions.  We shall begin by considering the axion ${\cal
A}_0^{(12)}$.  Certain field redefinitions are required in order to recast
the massless $D=8$ theory into a form in which the field ${\cal A}_0^{(12)}$
is covered by a derivative everywhere, namely 
\be
A_1^{(23)}\longrightarrow A_1^{(23)} + {\cal A}_0^{(12)} A_1^{(13)}\ ,
\qquad A_2^{(2)} \longrightarrow A_2^{(2)} + {\cal A}_0^{(12)} A_2^{(1)} 
\ .
\ee
One can easily check that these redefinitions, whose form is dictated by
requiring that ${\cal A}_0^{(12)}$ be covered by a derivative everywhere in
the bilinear Chern-Simons modifications displayed in (\ref{bilin8}), are
also sufficient to ensure that it is covered by a derivative in all of the
higher-order terms too. 

     In a similar manner, if instead we wish to use ${\cal A}_0^{(13)}$ for
the \ss reduction, we may ensure that it is covered by derivatives
everywhere by performing instead the following field
redefinitions\footnote{In fact, as we shall discuss later, it is possible to
cover more than one axion with derivatives {\it simultaneously}, thereby
permitting more exotic kinds of \ss reductions.} 
\be
A_1^{(23)}\longrightarrow A_1^{(23)} - {\cal A}_0^{(13)} A_1^{(12)}\ ,
\qquad A_2^{(3)} \longrightarrow A_2^{(3)} + {\cal A}_0^{(13)} A_2^{(1)}
\ .
\ee
The third of the axions of the ${\cal A}_0^{(ij)}$ type, namely
${\cal A}_0^{(23)}$, can be covered by a derivative everywhere if we instead
redefine fields according to:
\bea
{\cal A}_0^{(13)}& \longrightarrow& {\cal A}_0^{(13)} + {\cal A}_0^{(12)}
{\cal A}_0^{(23)}\ ,\nonumber\\
A_1^{(13)}&\longrightarrow & A_1^{(13)} + {\cal A}_0^{(23)} A_1^{(12)}\ ,
\qquad A_2^{(3)} \longrightarrow A_2^{(3)} + {\cal A}_0^{(23)} A_2^{(2)}
\ .
\eea
Finally, we note that the fourth axion, $A_0^{(123)}$, already appears
covered by a derivative everywhere in the Chern-Simons modifications to the
field strengths.  It does, however, appear undifferentiated in the ${\cal
L}_{FFA}$ term, given in (\ref{ffaterms}).  This is easily taken care of by
performing an integration by parts on the term where this field appears in
the action.  Thus, we have seen that we can perform a \ss reduction on any
of the four axions of the $D=8$ supergravity theory. 

     For the first three axions, ${\cal A}_0^{(ij)}$, where there are no
contributions arising from the ${\cal L}_{FFA}$ term, the spectrum of
massive fields is determined in precisely the manner described in the
previous subsection.  If we use ${\cal A}_0^{(12)}$ for the \ss reduction,
we find that three 2-form field strengths become massive, corresponding to
the fields ${\cal A}_1^{(4)}$, ${\cal A}_1^{(2)}$ and $A_1^{(13)}$.  It
turns out that there is then a residual $SL(2,\R)$ symmetry of the scalar
manifold, under which the first of these fields is a singlet, while the
latter two form a doublet.  There is also a massive 3-form field strength,
corresponding to $A_2^{(1)}$, and a massive 1-form field strength
corresponding to ${\cal A}_0^{(23)}$.  Both of these are singlets under
$SL(2,\R)$.  The complete structure of massive and massless fields is
displayed in Table 1 below.  The fields enclosed in angle brackets in the
massive column indicate those that are absorbed by the corresponding fields
appearing directly above them.  The axion $A_0^{(123)}$ is enclosed in
square brackets, to indicate that it is the ``ignorable coordinate'' $\chi$,
appearing always differentiated, in the $SL(2,\R)/SO(2)$ scalar manifold. 
The associated dilatonic scalar in this manifold is the combination
$\varphi\equiv -\ft12 \vec a_{123} \cdot \vec\phi$. Note that $B_2$ is the
2-form potential for the dual of the 4-form field strength, {\it i.e.}\
$^*F_4=dB_2$.  The dilaton vector for the cosmological term $\ft12 m^2
e^{\vec c\cdot \vec \phi}$ in the $D=7$ massive theory is given by $\vec
c=\vec b_{124}$ in this case, where $\vec b_{124}$ is defined in
(\ref{dilatonvec}). 

\bigskip\bigskip
\centerline{
\begin{tabular}{|c|c|c|}\hline Dimension
&Massive & Massless \\ \hline\hline
Doublets &$({\cal A}_1^{(2)}, A_1^{(13)})$  &
$(A_2^{(4)}, B_2)$ \\
& $\langle{\cal A}_0^{(14)}, A_0^{(234)} \rangle $   &
$({\cal A}_1^{(3)}, A_1^{(12)})$  \\
&    &  $({\cal A}_0^{(24)},  A_0^{(134)})$ \\
&    &  $({\cal A}_0^{(34)},  A_0^{(124)})$  \\
\hline
Singlets & $A_2^{(1)}$   &   $A_2^{(2)}\ A_2^{(3)}$    \\
& $\langle \ A_1^{(24)}\ \rangle$
&   ${\cal A}_1^{(1)}\ A_1^{(14)} \ A_1^{(23)} \ A_1^{(34)}$     \\
& ${\cal A}_1^{(4)}$ & ${\cal A}_0^{(13)}\ [\ A_0^{(123)}\ ]$ \\
& $\langle \ {\cal A}_0^{(12)} \ \rangle $ &  \\
& ${\cal A}_0^{(23)}$ & \\ \hline
\end{tabular}}
\bigskip
\centerline{Table 1: $SL(2,\R)$ multiplets for the ${\cal A}_0^{(12)}$
reduction}
\bigskip

     If the axion ${\cal A}_0^{(13)}$ is used instead for the \ss reduction,
then we find that the fields $({\cal A}_1^{(4)}, A_1^{(12)}, {\cal
A}_1^{(3)})$ become massive, absorbing $\langle {\cal A}_0^{(13)},
A_0^{(234)}, {\cal A}_0^{(14)} \rangle $ in the process. We find that these
fields form triplets under a residual $SL(3,\R)$ symmetry of the scalar
manifold.  The $SL(3,\R)/SO(3)$ manifold has three axionic ignorable
coordinates $[{\cal A}_0^{(34)}, A_0^{(123)}, A_0^{(124)}]$, Their
associated dilatonic scalars are the two combinations $\varphi_1 = - \ft12
\vec b_{34}\cdot\vec\phi$ and $\varphi_2 = -\ft1{2\sqrt3} (\vec a_{123}+
\vec a_{124})\cdot\vec\phi$.  There is also a singlet massive field
$A_2^{(1)}$, which absorbs $A_1^{(34)}$. These, and the remaining massless
fields, are listed Table 2 below.  They also fall into triplets and singlets
under the $SL(3,\R)$ symmetry group. As above, in the massive column, the
fields enclosed in angle brackets are absorbed by those immediately above
them.  The dilaton vector for the cosmological term is given by $\vec c=\vec
b_{134}$ in this case. 

\bigskip\bigskip
\centerline{
\begin{tabular}{|c|c|c|}\hline Dimension
 &Massive & Massless \\ \hline\hline
 Triplets &$({\cal A}_1^{(4)}, A_1^{(12)}, {\cal A}_1^{(3)})$  &  
  $(A_2^{(3)}, A_2^{(4)}, B_2)$ \\
          & $\langle{\cal A}_0^{(13)}, A_0^{(234)}, {\cal A}_0^{(14)}
	 \rangle $   &      
 $({\cal A}_1^{(1)}, A_1^{(23)}, A_1^{(24)})$  \\
	  &    &  $({\cal A}_1^{(2)}, A_1^{(13)}, A_1^{(14)})$     \\
	  &    &  $({\cal A}_0^{(23)}, {\cal A}_0^{(24)}, A_0^{(134)})$ \\
	  &    &  $[{\cal A}_0^{(34)},  A_0^{(123)}, A_0^{(124)}]$  \\ 
	  \hline
 Singlets & $A_2^{(1)}$   &   $A_2^{(2)}$    \\
	  & $\langle \ A_1^{(34)}\ \rangle$   
	  &   ${\cal A}_0^{(12)}$     \\ \hline
 \end{tabular}}
 \bigskip
 \centerline{Table 2: $SL(3,\R)$ multiplets for the ${\cal A}_0^{(13)}$
 reduction}
 \bigskip

     Turning now to the \ss reduction using the third of the ${\cal
A}_0^{(ij)}$ axions, namely ${\cal A}_0^{(23)}$, we find once again there is
an $SL(3,\R)$ symmetry of the scalar manifold, with $[{\cal A}_0^{(34)},
A_0^{(123)}, A_0^{(124)}]$ as the ignorable coordinates. Their associated
dilatonic scalars are the two combinations $\varphi_1=-\ft12 \vec
b_{23}\cdot \vec\phi$ and $\varphi_2= - \ft1{2\sqrt3} (\vec a_{123}+ \vec
a_{124})\cdot\vec\phi$.  The triplets and singlets under $SL(3,\R)$ are
listed in Table 3 below, using the same notation as in the previous cases. 
The dilaton vector for the cosmological term is $\vec c=\vec b_{234}$.  Note
that, although the surviving $SL(3,\R)$ symmetry group for this case is the
same as for the \ss reduction using ${\cal A}_0^{(13)}$, the two $D=7$
theories are quite distinct; in particular, there is a massive axion in the
${\cal A}_0^{(23)}$ reduction, while in the ${\cal A}_0^{(13)}$ reduction
the analogous axion is massless. 

\bigskip\bigskip
\centerline{
\begin{tabular}{|c|c|c|}\hline Dimension
&Massive & Massless \\ \hline\hline
Triplets &$({\cal A}_1^{(3)}, {\cal A}_1^{(4)}, A_1^{(12)})$  &
$(A_2^{(3)}, A_2^{(4)}, B_2)$ \\
& $\langle{\cal A}_0^{(23)}, {\cal A}_0^{(24)}, A_0^{(134)} \rangle $   &
$({\cal A}_1^{(1)}, A_1^{(23)}, A_1^{(24)})$  \\
&    &  $({\cal A}_1^{(2)}, A_1^{(13)}, A_1^{(14)})$     \\
&    &  $({\cal A}_0^{(13)}, {\cal A}_0^{(14)}, A_0^{(234)})$ \\
&    &  $[{\cal A}_0^{(34)},  A_0^{(123)}, A_0^{(124)}]$  \\
\hline
Singlets & $A_2^{(2)}$   &   $A_2^{(1)}$    \\
& $\langle \ A_1^{(34)}\ \rangle$
&       \\ 
& ${\cal A}_0^{(12)}$ &\\ \hline
\end{tabular}}
\bigskip
\centerline{Table 3: $SL(3,\R)$ multiplets for the ${\cal A}_0^{(23)}$
reduction}
\bigskip

     Finally, we turn to the \ss reduction using the axion $A_0^{(123)}$
derived from the 3-form potential occurring in $D=11$. In this case, we find
that there is a residual $SL(4,\R)$ symmetry of the scalar manifold, with
all of the axions $[{\cal A}_0^{(ij)}]$ being ignorable coordinates.  The
associated dilatonic coordinates are the three combinations of the
$\vec\phi$ fields that are orthogonal to the dilaton vector for the
cosmological term, which is given by $\vec c=\vec a_{1234}$ in this case. 
The various fields fall into sextets, quartets and singlets under
$SL(4,\R)$, as indicated in Table 4 below. Note that, in this case, the
field $A_3$ becomes massive, not by the usual mechanism of absorbing a
2-form potential, but instead by acquiring a topological mass term in the
$D=7$ Lagrangian.  This arises because the term $-\ft12 \td F_4\wedge \td
F_4 A_0^{(123)}$ in the $D=8$ ${\cal L}_{FFA}$ term gives $-\ft12 \td
F_4\wedge A_3\wedge dA_0^{(123)}$ after integration by parts, and this then
reduces, on imposing the \ss reduction ansatz $A_0^{(123)}(x,z)\rightarrow
mz+ A_0^{(123)}(x)$, to give a term 
\be
-\ft12 m F_4\wedge A_3 \label{topmass}
\ee
in the $D=7$ Lagrangian.  The supergravity theory arising in this case is
the maximal $N=2$ extension of the $N=1$ theory with a topological mass term
found in \cite{mtn} (with the Yang-Mills gauge coupling constant set equal
to zero). 

\bigskip\bigskip
\centerline{
\begin{tabular}{|c|c|c|}\hline Dimension
&Massive & Massless \\ \hline\hline
Sextets & & $(A_1^{(12)}, A_1^{(13)}, A_1^{(14)}, A_1^{(23)},
A_1^{(24)}, A_1^{(34)} )$ \\
& & $[{\cal A}_0^{(12)}, {\cal A}_0^{(13)}, {\cal A}_0^{(14)}, 
{\cal A}_0^{(23)}, {\cal A}_0^{(24)}, A_0^{(34)}]$ \\ \hline
Quadruplets &$({\cal A}_1^{(1)}, {\cal A}_1^{(2)}, {\cal A}_1^{(3)},
{\cal A}_1^{(4)})$  &
$(A_2^{(1)}, A_2^{(2)}, A_2^{(3)}, A_2^{(4)})$ \\
& $\langle A_0^{(234)}, A_0^{(134)}, A_0^{(124)}, A_0^{(123)} \rangle $   &
 \\
\hline
Singlets & $A_3$  &      \\ \hline
\end{tabular}}
\bigskip
\centerline{Table 4: $SL(4,\R)$ multiplets for the $A_0^{(123)}$
reduction}
\bigskip

     We have seen above that, depending on the choice of which of the four
axions of the $D=8$ theory is used for the \ss reduction step, we can obtain
four inequivalent massive $D=7$ maximal supergravity theories.  It is also
easy to see that we get two further inequivalent theories arising in $D=7$
by performing ordinary \kk reductions of the massive IIA theory in D=10, and
of the massive theory in $D=8$ that was discussed in the previous
subsection.  The simplest way of seeing the inequivalence of these is by
observing that the massive IIA theory, for which just a single 2-form
potential is massive in $D=10$, will give a theory in $D=7$ that has one
massive 2-form potential, three massive 1-forms, and three massive axions in
$D=7$.  On the other hand, the massive $D=8$ theory constructed in the
previous subsection has a massive 2-form potential and two massive 1-forms
in $D=8$, yielding, after ordinary \kk reduction, a massive 2-form
potential, three massive 1-forms, and two massive axions in $D=7$. These two
theories thus differ in their axionic sectors from the four
already-inequivalent theories constructed above. Although these latter have
the same counting of massive 2-form and 1-form potentials, they have either
no massive axion, or one massive axion. 

     It is of interest also to examine the surviving symmetries of the
scalar Lagrangians in $D=7$ for the two additional theories just discussed,
obtained either from the dimensional reduction of the massive IIA theory in
$D=10$, or from the new massive theory in $D=8$ that we obtained in the
previous subsection.  We shall not present the details here, but simply
record that the first of these two theories has a surviving $SL(3,\R)$
symmetry in $D=7$, while the second has an $SL(2,\R)$ symmetry.  Finally, we
remark that, as in the case of the massive $D=8$ supergravity constructed in
the previous subsection, one can also seek in all six of the $D=7$ theories
for additional abelian symmetries, associated either with shift symmetries
of the axions that do not participate as ignorable coordinates in the
non-abelian coset manifolds as already considered, or else associated with
shift symmetries of dilatonic scalars, together with appropriate rescalings
of the gauge potentials.  Abelian symmetries of the former type are somewhat
involved to analyse in general, since one needs to determine how many of the
``surplus'' axions admit shift symmetries (related issues are discussed in
the next section).  Extra abelian symmetries of the second type are easier
to classify.  We find that each combination of dilatonic scalars that is
orthogonal to the dilaton vector for the cosmological term, and that is also
orthogonal to all of the dilaton vectors for the axions involved in the
non-abelian part of the scalar manifold, allows such an independent abelian
symmetry. This implies that the non-abelian symmetries $SL(2,\R)$,
$SL(3,\R)$ and $SL(4,\R)$ which we found for the various $D=7$ theories
become (at least) $SL(2,\R)\times \R\times \R$, $SL(3,\R)\times \R$, and
$SL(4,\R)$ respectively, when the dilatonic shift symmetries are included. 

      As we proceed to lower dimensions, the numbers of inequivalent massive
theories proliferates.  The numbers of axions in $D=7,6,5$ and 4 are 10, 20,
36 and 63 respectively.\footnote{In $D=5$ we are including $A_3$ in the
counting of axions, since its field strength $F_4$ can be dualised to a
1-form field strength for another axion.  Similarly, in $D=4$ we are
including the 2-form potentials $A_2^{(i)}$.  As we shall discuss further in
the conclusions, for the purposes of constructing consistent truncations of
$D=11$ to give lower-dimensional theories, it is better to leave these
fields in their original undualised formulations, in which case they would
reduce by the ordinary \kk procedure rather than \ss procedure.} In each
dimension, each axion, if used for a \ss reduction of the kind we are
considering here, will give rise to a different massive theory in one lower
dimension.  Furthermore, we accumulate additional inequivalent theories from
the ordinary \kk reductions of higher-dimensional massive theories.  Thus,
the total numbers of massive theories of the kind we are constructing here 
will be 15, 35, 71 and 134 in $D=6, 5, 4$ and 3, respectively.  In addition,
there is one more massive theory in each dimension that comes from the \kk
reduction of the massive IIA theory in $D=10$. The massive theories may
admit consistent truncations to theories with less than maximal
supersymmetry.  For example in $D=4$, some of the 72 massive $N=8$ theories
may admit truncations to $N=1$, in which case they may reduce to the models
considered in \cite{cgr}, for which domain-wall solutions were given.

\section{\ss reduction using multiple axions}

     In the examples of \ss reduction that we have considered so far, we
have imposed the standard \kk ansatz on all of the higher-dimensional
fields, with the exception of a single axion, for which the more general
ansatz (\ref{ssans}) was allowed.  As we discussed above, the ability to
perform such a consistent truncation depends upon the fact that there exists
a choice of field variables in which the chosen axion $\chi$ is covered by a
derivative everywhere in the the higher-dimensional Lagrangian.  Although we
did not exploit it in the previous section, we also remarked that one can in
fact typically choose field variables in the higher dimension such that more
than one of the axions are simultaneously covered by derivatives everywhere.
 Under these circumstances, it is then possible to perform a family of \ss
reductions in which all the derivative-covered axions satisfy ans\"atze of
the form (\ref{ssans}), with independent mass parameters for each.  We shall
explore below the structure of the resulting lower-dimensional theories that
are obtained in such cases. The counting of the number of axions that can
simultaneously be covered with derivatives is a somewhat involved one in
general, and we shall not attempt a complete discussion.  Clearly in
principle it can be established, dimension by dimension, by enumerating all
of the cases.  Rather than do this, we shall make some general observations,
and then shall consider the example of $D=8$ supergravity in detail, to
illustrate the possibilities. 

     From the general form of the Lagrangian for maximal supergravity in $D$
dimensions, as given in the Appendix, it is clear that all of the axions of
the form $A_0^{(ijk)}$ can easily be simultaneously covered by derivatives,
without the need for any field redefinitions. They already appear in the
desired form in the Chern-Simons modifications (\ref{A.6}), while in the
${\cal L}_{FFA}$ terms (\ref{ffaterms}) it is a simple matter of performing
appropriate integrations by parts in order to cover them with derivatives
there too.  In addition, it is possible that appropriate field redefinitions
can allow one or more of the ${\cal A}_0^{(ij)}$ axions to be simultaneously
covered as well.  The enumeration of all the possibilities is complicated,
however, since it may well be, for example, that more axions of the ${\cal
A}_0^{(ij)}$ type can be covered if fewer of the $A_0^{(ijk)}$ axions are
covered, and so on. 

     Let us now consider the specific example of $D=8$ supergravity, which
contains four axions; ${\cal A}_0^{(12)}$, ${\cal A}_0^{(13)}$, ${\cal
A}_0^{(23)}$ and $A_0^{(123)}$.  It is clear from the expression for ${\cal
F}_1^{(13)}$ in (\ref{bilin8}) that it is not possible to put derivatives on
${\cal A}_0^{(12)}$ and ${\cal A}_0^{(23)}$ simultaneously. We can, however,
cover three out of the four axions, for example ${\cal A}_0^{(12)}$, ${\cal
A}_0^{(13)}$ and $A_0^{(123)}$.  This is achieved by making the field
redefinitions 
\bea
A_1^{(23)} &\longrightarrow & A_1^{(23)} - {\cal A}_0^{(13)} A_1^{(12)} 
+ {\cal A}_0^{(12)} A_1^{(13)} \ ,\nonumber\\
A_2^{(2)} &\longrightarrow & A_2^{(2)} + {\cal A}_0^{(12)} A_2^{(1)}\ ,
\qquad
A_2^{(3)} \longrightarrow  A_2^{(3)} + {\cal A}_0^{(13)} A_2^{(1)}\ ,
\label{3ax}
\eea
together with the necessary intergration by parts in the ${\cal L}_{FFA}$
term.

     We shall now discuss the general structure of the $D=7$ theory that
results from \ss reduction of the $D=8$ theory, after the redefinitions
(\ref{3ax}), with the ans\"atze 
\bea
{\cal A}_0^{(12)}(x,z)&\longrightarrow & m_1\,  z + {\cal A}_0^{(12)}(x)\ ,
\nonumber\\
{\cal A}_0^{(13)}(x,z)&\longrightarrow & m_2\,  z + {\cal A}_0^{(13)}(x)\ ,
\label{axans}\\
A_0^{(123)}(x,z) &\longrightarrow & m_3\,  z + A_0^{(123)}(x)\ .\nonumber
\eea
We find that the form of the resulting theory in $D=7$ is
\bea
e^{-1}{\cal L}&=& R -\ft12 (\del\vec\phi)^2 
-\ft12 e^{{\vec b}_{12}\cdot\vec\phi} 
(\del {\cal A}_0^{(12)} - m_1 {\cal A}_1^{(4)})^2
\nonumber\\
&&
-\ft12 e^{{\vec b}_{13}\cdot\vec\phi}
(\del {\cal A}_0^{(13)} - m_2 {\cal A}_1^{(4)}-{\cal A}_0^{(23)} 
     \del {\cal A}_0^{(12)} +m_1 {\cal A}_0^{(23)} {\cal A}_1^{(4)})^2 
\nonumber\\
&&
-\ft12 e^{{\vec b}_{14}\cdot\vec\phi} 
(\del {\cal A}_0^{(14)} + m_1 {\cal A}_1^{(2)}+ m_2 {\cal A}_1^{(3)})^2
-\ft12 e^{{\vec b}_{23}\cdot\vec\phi} (\del {\cal A}_0^{(23)})^2
-\ft12 e^{{\vec b}_{24}\cdot\vec\phi} (\del {\cal A}_0^{(24)})^2 
\nonumber\\
&&
-\ft12 e^{{\vec b}_{34}\cdot\vec\phi} (\del {\cal A}_0^{(34)})^2
-\ft12 e^{{\vec a}_{123}\cdot\vec\phi} 
(\del A_0^{(123)} - m_3 {\cal A}_1^{(4)})^2
\nonumber\\
&&
-\ft12 e^{{\vec a}_{124}\cdot\vec\phi}
(\del A_0^{(124)} + m_3 {\cal A}_1^{(3)})^2
-\ft12 e^{{\vec a}_{134}\cdot\vec\phi}
(\del A_0^{(134)} - m_3 {\cal A}_1^{(2)})^2
\label{d73axlag}\\
&&
-\ft12 e^{{\vec a}_{234}\cdot\vec\phi}
(\del A_0^{(234)} + m_3 {\cal A}_1^{(1)} +m_1 A_1^{(13)} -m_2 A_1^{(12)})^2 
-\ft14 \sum_{i=1}^4 e^{{\vec b}_i\cdot\vec\phi} ({\cal F}_2^{(i)})^2
\nonumber\\
&&
-\ft14 \sum_{i<j\le 3} e^{{\vec a}_{ij}\cdot\vec\phi} (F_2^{(ij)})^2
-\ft14 e^{{\vec a}_{14}\cdot\vec\phi} (F_2^{(14)})^2
-\ft14 e^{{\vec a}_{24}\cdot\vec\phi} (F_2^{(24)} +m_1 A_2^{(1)})^2
\nonumber\\
&&
-\ft14 e^{{\vec a}_{34}\cdot\vec\phi} (F_2^{(34)} +m_2 A_2^{(1)})^2
-\ft1{12} \sum_{i=1}^4 e^{{\vec a}_i\cdot\vec\phi} (F_3^{(i)})^2
-\ft1{48} e^{\vec a\cdot\vec\phi} F_4^2 -\ft12 m_3 F_4\wedge A_3 
\nonumber\\
&&
-\ft12 m_1^2\,  e^{\vec b_{124}\cdot \vec\phi}
-\ft12 m_3^2\,  e^{\vec a_{1234}\cdot \vec\phi}
-\ft12 (m_2 -m_1 {\cal A}_0^{(23)})^2 e^{\vec b_{134}\cdot \vec\phi}
+ \cdots
\ ,\nonumber
\eea
where we have suppressed certain higher-order terms that are not of
relevance for the present discussion. 

     We see from the final line of the Lagrangian that there are two
standard-type cosmological terms, namely those with coefficients $m_1^2$ and
$m_3^2$, together with the final term, which requires some further
discussion. If $m_1$ is non-zero, we can make the field redefinition 
\be
{\cal A}_0^{(23)} \longrightarrow {\cal A}_0^{(23)} +\fft{m_2}{m_1}\ ,
\ee
in which case the final term becomes a standard mass term for ${\cal
A}_0^{(23)}$.  Note also that, at the same time, the $m_2 {\cal A}_1^{(4)}$
term in the kinetic term $(\del{\cal A}_0^{(13)} +\cdots )^2$ is removed. 
On the other hand, if $m_1=0$, then the final term in the Lagrangian
(\ref{d73axlag}) becomes a standard cosmological term, while the first
cosmological term, $-\ft12 m_1^2\, e^{{\vec b}_{124} \cdot\vec\phi}$,
disappears.  Thus, either way, there are two cosmological-type terms in the
$D=7$ Lagrangian. 
    
     In lower dimensions, the possibilities for performing simultaneous \ss
reductions on multiple axions become more numerous. For example, we have
found by explicit computation that in all dimensions, all the axions of the
form ${\cal A}_0^{(1a)}$ and $A_0^{(abc)}$ can simultaneously be covered by
derivatives, where the indices $a,b,\ldots$ run over the values 2 to $11-D$.
 It is interesting that these are precisely the Ramond-Ramond axions, from
the viewpoint of the type IIA string \cite{lpsweyl}.  In $D=5$, there is
another field that describes an axionic-type degree of freedom, namely
$A_3$, whose field strength $F_4$ is dual to another 1-form field strength. 
The field $A_3$ is in the Ramond-Ramond sector, and we find that in $D=5$ it
too can be covered by derivatives at the same time as the RR axions ${\cal
A}_0^{(1a)}$ and $A_0^{(abc)}$. Similarly, in $D=4$ where the potentials
$A_2^{(i)}$ describe axionic-type degrees of freedom, we find that the
Ramond-Ramond subset $A_2^{(a)}$ can be covered by derivatives at the same
time as the RR axions ${\cal A}_0^{(1a)}$ and $A_0^{(abc)}$.  Thus in total
we have $2^{9-D}$ RR fields in $D$ dimensions that can be simultaneously
covered by derivatives. 
 
     Other sets of axions may also be covered by derivatives simultaneously.
 For example, we have already seen in this section that in $D=8$, we may
cover three of the four axions simultaneously.  These comprise the two
Ramond-Ramond axions ${\cal A}_0^{(12)}$ and ${\cal A}_0^{(12)}$, together
with the NS-NS axion $A_0^{(123)}$.  Another example, in $D=4$, is the set
of all 35 axions  $A_0^{(ijk)}$, which  can be covered by derivatives
simultaneously.  These comprise 20 Ramond-Ramond fields $A_0^{(abc)}$ and 15
NS-NS fields $A_0^{(1ab)}$, and their total number exceeds the total of 32
Ramond-Ramond axions described above. Using all the $A_0^{(ijk)}$ axions, a
\ss reduction to $D=3$ can be performed that involves 35 independent mass
parameters. However, this does not imply that there will be 35 different
cosmological terms in $D=3$, just as we saw above in $D=7$ that there are 2
cosmological terms rather than 3.  In $D=7$, the reason for this can be
identified by looking at the final term in (\ref{d73axlag}); this indeed
would be a cosmological term were it not for the appearance of the ${\cal
A}_0^{(23)}$ field there, as we have seen. The presence of this term can be
traced back to the Chern-Simons modification structure for the ${\cal
F}_1^{(13)}$ field strength in (\ref{bilin8}).  In fact, Chern-Simons
modifications of this type are responsible for another feature in
supergravity theories; namely that they govern the structure of the allowed
combinations of 1-form field strengths that can participate simultaneously
in $(D-3)$-brane solutions to the massless theories.  The maximum number
$N_c$ of cosmological terms in massive $D$-dimensional supergravity is the
same as the maximum number of 1-forms that can participate in
``$(D-3)$-brane'' solutions in $(D+1)$ dimensions, and thus from results
obtained, for example, in \cite{lpsol}, we find that $N_c=1$ in $D=8$;
$N_c=2$ in $D=7$ and $D=6$; $N_c=4$ in $D=5$ and $D=4$; and $N_c=7$ in
$D=3$. 

\section{Domain wall solutions}

     In this section, we turn to a consideration of the $(D-2)$-brane
solutions that arise in massive supergravity theories.  As we discussed in
the introduction, these are the natural endpoints of vertical dimensional
reduction, and, in fact, we have seen that they {\it require} that the usual
\kk reduction procedure be generalised to the \ss reduction.  In the
dimensional reduction of the higher-dimensional supergravity theory, the \ss
reduction is responsible for generating the cosmological-type term occurring
in the lower-dimensional Lagrangian that supports the $(D-2)$-brane
solution. Thus the general form of the relevant part of the
lower-dimensional Lagrangian is \cite{lpss2} 
\be
{\cal L}= e R -\ft12 e (\del \phi)^2 -\ft12  m^2 e e^{a \phi}\ .\label{blag}
\ee
It is useful \cite{lpss1} to re-express the parameter $a$ describing
the dilaton coupling in terms of the quantity $\Delta$, defined by
\be
a^2=\Delta + \fft{2(D-1)}{(D-2)}\ ,\label{delta}
\ee
where $D$ denotes the spacetime dimension of the lower-dimensional theory.
The equations of motion following from the Lagrangian (\ref{blag}) admit
$(D-2)$-brane solutions of the form \cite{lpss1,lpss2}
\bea
ds^2 &=& H^{\ft4{\Delta(D-2)}} \, \eta_{\mu\nu}\, dx^\mu dx^\nu 
+ H^{\ft{4(D-1)}{\Delta(D-2)}} \, dy^2 \ ,\label{dwsol}\\
e^\phi &=& H^{2a/\Delta} \ ,\nonumber
\eea
where $H$ is an harmonic function on the 1-dimensional transverse space with
coordinate $y$, of the general form $H = c \pm M y$, where $c$ is an
arbitrary constant, and $M= \ft12 m \sqrt\Delta$.  The curvature of the
metric tends to zero at large values of $|y|$; it diverges if $H$ tends to
zero, which can be avoided by taking $H$ to be of the form $H=c + M |y|$,
with the constant $c$ being positive. There is then just a delta-function
singularity in the curvature at $y=0$, corresponding to a discontinuity in
the gradient of the ``confining potential.''  The solution so obtained
describes a domain wall. (The solution (\ref{dwsol}) in $D=4$ was also
obtained in \cite{c}.  For a recent review of domain-walls in $D=4$
supergravities, see \cite{cs}.) 

     The simplest solutions of this kind occurring in the massive
supergravity theories we have constructed in this paper arise when just a
single axion is used in the \ss reduction procedure.  In such a case, the
field $\phi$ in (\ref{blag}) is just the linear combination of all the
dilatonic fields $\vec\phi$ that lies in the direction parallel to the
dilaton vector $\vec c$ of the cosmological term $-\ft12 m^2 e^{\vec c\cdot
\vec\phi}$ in the \ss reduced theory.  It is easily verified that all the
possible dilaton vectors, given by $\vec a_{ijk\ell}$ or $\vec b_{ijk}$ in
(\ref{dilatonvec}), have the property that they have magnitude
$\sqrt{2(3D-5)/(D-2)}$ and hence, after properly normalising $\phi$ so that
it has a canonical kinetic term, we find that the value of the dilaton
coupling constant $a$ in (\ref{blag}) corresponds to $\Delta=4$.  Thus, in
all these cases, the form of the harmonic function $H$ is $H=c \pm m y$,
where $m$ is the same as the mass parameter used in the \ss ansatz
(\ref{ssans}). 

     As we have discussed in the previous section, it is possible to obtain
massive supergravity theories in $D\le 7$ that have two or more cosmological
terms with independent mass parameters, by performing a more general form of
\ss reduction using more than one axion.  Just as this can be done for the
$p$-brane solutions supported by charges carried by field strengths of rank
$>0$, one can make a consistent truncation of such a theory with multiple
cosmological terms, thereby arriving at a Lagrangian that is again of the
form (\ref{blag}), but with a value of the dilaton coupling $a$ that no
longer corresponds to $\Delta=4$. In fact, it is not hard to see that if the
massive supergravity theory has $N_c$ cosmological terms, then consistent
truncations to the form (\ref{blag}) with $\Delta=4/N$ are possible, where
$1\le N\le N_c$. For example, in $D=7$ we obtain the massive theory
described by (\ref{d73axlag}), with two cosmological terms.  As well as the
$\Delta=4$ domain-wall solutions obtainable by using either one or the other
of these, we can also construct a $\Delta=2$ domain-wall solution.  The
necessary truncation to a single scalar $\phi$ is achieved, as in
\cite{lpsol}, by introducing a constant unit vector $\vec n$, defining
$\vec\phi = \vec n \phi + \vec\phi_\perp$ where $\vec
n\cdot\vec\phi_\perp=0$, and requiring that the truncation
$\vec\phi_\perp=0$ be consistent with the equations of motion for
$\vec\phi_\perp$.  The resulting linear algebra implies, for a truncation
where the masses $m_1$ and $m_3$ are both non-zero, that $m_1^2 \, \vec
b_{124} +m_3^2 \, \vec a_{1234} = a \, \vec n\, (m_1^2 +m_3^2)$, and hence
$m_1^2=m_3^2$ and $a^2=\ft{22}5$.  From (\ref{delta}), it therefore follows
that $\Delta=2$. 

     It is straightforward to see that the $\Delta=2$ domain wall solution
in $D=7$ that we have just constructed can itself be obtained from the
appropriate 5-brane solution with $\Delta=2$ in $D=8$, by vertical
dimensional reduction. This is just like the procedure for vertical
reduction that we described in the introduction, where now the field
strength in $D=8$ that supports the 5-brane solution is obtained by setting
equal ${\cal F}_1^{(12)}$ and $F_1^{(123)}$, together with the corresponding
truncation of dilatonic scalars. In a similar manner, we may construct
domain wall solutions with $\Delta=4/N$ in any lower dimension too, for any
integer $N\le N_c$, where $N_c$ is the number of independent cosmological
terms in that dimension.  Once again, these solutions can all be interpreted
as the vertical reductions of $(D-3)$-branes in one higher dimension, with
the same values of $\Delta$.\footnote{There also exist more complicated
$(D-3)$-brane solutions with $\Delta=24/(N(N+1)(N+2))$, which arise as
solutions of the $SL(N+1,\R)$ Toda equations \cite{lptoda}.  These are
non-supersymmetric, and we shall not discuss their possible vertical
reduction to domain walls here.}  These domain walls will preserve the same
fraction of the supersymmetry as their $(D-3)$-brane oxidations, since
dimensional reduction of the kind we are considering in this paper, whether
of the ordinary \kk sort or of the more general \ss sort, preserves all of
the unbroken components of supersymmetry in the corresponding supergravity
theories. Thus, solutions with $N=1$, $N=2$ and $N=3$ preserve $2^{-N}$ of
the supersymmetry, while solutions with $N\ge4$ all preserve $1/16$ of the
supersymmetry \cite{lpsol}. 

      We may also construct more general kinds of domain-wall solutions,
precisely analogous to those that can be built for other kinds of
$p$-branes.  One such class of solution is the multi-scalar generalisations
of the solutions with $\Delta=4/N$ with $N\ge 2$.  Such solutions exist for
$p$-branes of arbitrary dimension \cite{lpmult}, and the domain walls that
we are considering in this paper are no exception.  They arise by relaxing
the constraint in the single-scalar solutions that the charges carried by
all participating field strengths be equal.  The greater generality of these
solutions is achieved by having more than one independent combination of the
dilatonic scalars be non-vanishing. In the context of domain-wall solutions,
the independence of the charges translates into an independence of the mass
parameters appearing as coefficients of the cosmological terms in the
corresponding massive supergravity theories.  For example, we may construct
a two-scalar generalisation of the $\Delta=2$ domain wall in $D=7$ discussed
above, in which the parameters $m_1$ and $m_3$ are no longer equal.  In
general, for a massive supergravity with cosmological terms of the form 
\be
{\cal L}_{\rm cosmo} = -\ft12 \sum_{\a=1}^N m_\a^2\, e^{\vec c_\a \cdot
\vec\phi} \ ,\label{gencos}
\ee
multi-scalar solutions exist if the dilaton vectors satisfy $\vec c_\a
\cdot\vec c_\b = 4\delta_{\a\b} +2(D-1)/(D-2)$; this is precisely the
relation that holds for the $\Delta=4/N$ single-scalar solutions that we
discussed previously.  Their multi-scalar generalisations take the form
\cite{lpmult} 
\bea
ds^2 &=& \prod_{\a=1}^N H_\a^{1/(D-2)}\, \eta_{\mu\nu} dx^\mu dx^\nu +
       \prod_{\a=1}^N H_\a^{(D-1)/(D-2)}\, dy^2 \ ,\nonumber\\
e^{\ft12\varphi_\a} &=& H_\a\, \prod_{\b=1}^N H_\b^{(D-1)/(D-2)}\ ,
\label{dwmult}
\eea
where $\varphi_\a \equiv \vec c_\a \cdot \vec\phi$, and $H_\a =1 + m_\a
|y|$.  (We have chosen the constant terms for all the harmonic functions to
equal unity, for simplicity.  More generally, we could allow arbitrary
constants for each harmonic function.)  These multi-scalar domain walls
reduce to single-scalar solutions with $\Delta=4/N$ if the mass parameters
$m_\a$ are all set to be equal. 

     Another generalisation of the domain-wall solutions that can
immediately be made is to consider the case of non-extremal domain walls. 
It was shown in \cite{dlp} that there is a universal prescription, for
$p$-brane solutions of all dimensions, for constructing non-extremal black
$p$-branes from the corresponding extremal ones.  When this is applied to
the case of domain walls, the general multi-scalar black solutions are given
by 
\bea
ds^2 &=& \prod_{\a=1}^N H_\a^{1/(D-2)}( -e^{2f} dt^2 + dx^i dx^i) +
       e^{-2f}\prod_{\a=1}^N H_\a^{(D-1)/(D-2)}\, dy^2 \ ,\nonumber\\
       e^{\ft12\varphi_\a} &=& H_\a\, \prod_{\b=1}^N H_\b^{(D-1)/(D-2)}
       \ ,
\label{dwblack}
\eea
where the harmonic functions are now given by
\be
H_\a = 1 + k |y| \sinh^2\mu_\a \ ,
\ee
and the function $f$ is given by
\be
e^{2f} = 1 - k |y| \ .
\ee
The previous extremal solutions are recovered in the appropriate limit,
where the parameter $k$ is sent to zero at the same time as one sends the
parameters $\mu_\a$ to infinity.  The single-scalar black solutions are
recovered if, instead, the parameters $\mu_a$ are set to be equal. 

\section{Domain walls as higher-dimensional solutions}

     We saw in the previous section that there is a correspondence between
the structure of cosmological terms in a massive supergravity in $(D-1)$
dimensions obtained by \ss reduction from $D$ dimensions, and the structures
of the possible consistent truncations in the 1-form field strength sector
of the massless supergravity in $D$ dimensions.  This correspondence in fact
brings us back to the discussion given in the Introduction that motivated
this work. This correspondence arises precisely because any $(D-3)$-brane
solution of massless supergravity in $D$ dimensions can be vertically
reduced to a domain-wall solution in $(D-1)$ dimensions with, as we saw in
the Introduction, a field configuration that forces us into a \ss reduction
scheme.  Conversely, any domain-wall solution to one of the massive
supergravities can be oxidised back to its progenitor solution in the
massless supergravity whence the massive theory came. 

     When we make such an oxidation, we are taking advantage of the fact
that all the dimensional-reduction schemes considered in this paper involve
{\it consistent truncations}. Thus solutions to lower-dimensional theories
may simply be {\it reinterpreted} as solutions to the higher-dimensional
theories whence the lower-dimensional theories arise by reduction. From a
domain-wall solution, one thus obtains by oxidation a stack of
``$(D-3)$-branes'' in $(D+1)$ dimensions, {\it i.e.}\ a solution with the
same worldvolume dimension as the starting domain wall, but with an
additional transverse dimension. Note that this oxidation process does not
reverse the stacking-up procedure itself and does not therefore extract a
single isotropic $(D-3)$-brane from the stack. To do so would change the
form of the harmonic function $H(y)$, and this does not happen under
dimensional oxidation. Note, however, that the value of $\Delta$ is
nonetheless preserved under dimensional reduction and oxidation
\cite{lpss1,lpsvert}. 

     Thus far, we have been discussing the single-step oxidation from a
domain wall in a particular dimension to a stack of $(D-3)$-branes in one
dimension higher.  Of course, one can continue the oxidation process further
on up.  Since the dimensional reduction of supergravity theories is a
consistent procedure both for the standard \kk type of reduction and also
for the \ss type of reduction, it follows that all solutions of a given
dimensionally-reduced theory can be traced back to the highest dimension in
which that theory has its origins. In this paper, our principal focus has
been on the massive theories originating from $D=11$ supergravity, and thus
we may consider the oxidation of all lower-dimensional domain-wall solutions
up to $D=11$.  Of course, this only makes sense for those domain walls in
lower dimensions that are solutions of the massive supergravities derivable
from $D=11$.  In particular, we have seen that the dimensional reduction of
the massive IIA theory in $D=10$ is inequivalent to all massive theories
that can be obtained by reduction from $D=11$, and thus there is no sense in
which solutions of the massive IIA theory can be related to those of $D=11$
supergravity. 

      The 6-brane solution of the massive supergravity in $D=8$ that we
constructed in section 2.1 has the metric $ds^2_8= H^{1/6} dx^\mu dx_\mu +
H^{7/6} dy^2$.  This oxidises to $ds^2_9= dx^\mu dx_\mu + H (dy^2 +dz_3^2)$
in $D=9$, and then to $ds^2_{10}= H^{-1/8} dx^\mu dx_\mu + H^{7/8} (dy^2+
dz_2^2 + dz_3^2)$ in $D=10$.  Finally, upon oxidation to $D=11$, it becomes 
\be
ds_{11}^2 = dx^\mu dx_\mu + H (dy^2 + dz_2^2 + dz_3^2) + H^{-1} (dz_1 + 
{\cal A}_1^{(1)})^2 \ ,\label{8dwox}
\ee
where ${\cal A}_1^{(1)}= m z_3\, dz_2$, and the coordinates $x^\mu$ span the
7-dimensional worldvolume of the solution. The harmonic function $H$ remains
the same under these oxidation steps, and continues to involve only the
original transverse coordinate $y$ as did the starting 6-brane in $D=8$.
Owing to the \ss oxidation step, the higher-dimensional solutions have a
dependence on $z_3$, either in the reduction axion in the $D=9$ case, or in
its $D=10$ progenitor ${\cal A}_1^{(1)}$, or finally in the $D=11$ metric,
as one sees in (\ref{8dwox}). All of these oxidised solutions have the same
worldvolume Poincar\'e symmetries as the original 6-brane in $D=8$, but
owing to their non-isotropic structure in the transverse space, they do not
have interpretations as standard $p$-branes. 

     In $D=7$, we have seen that there are five massive supergravity
theories that can be obtained from $D=11$ supergravity. One of these is the
diagonal dimensional reduction of the $D=8$ theory that we discussed in
section (2.1). Thus, the oxidation of its domain wall solution to $D=11$
gives the same metric as above, but with the 7$^{\rm th}$ worldvolume
coordinate replaced by the compactification coordinate $z_4$: 
\be
ds_{11}^2 = dx^\mu dx_\mu + dz_4^2+ H (dy^2 + dz_2^2 + dz_3^2) + H^{-1} (dz_1 +
{\cal A}_1^{(1)})^2 \ .\label{7dwox}
\ee
The vector potential is again given by ${\cal A}_1^{(1)}= m z_3\, dz_2$, and
the coordinates $x^\mu$ now span the 6-dimensional worldvolume of the
solution, which retains the $d=6$ Poincar\'e symmetry of the starting
solution.  The remaining four massive $D=7$ theories that come from $D=11$
are the ones that we constructed in section 2.2.  The corresponding four
$\Delta=4$ domain walls in $D=7$ oxidise in different fashions, eventually
giving the following metrics in $D=11$: 
\bea
{\cal A}_0^{(12)}:&& ds_{11}^2 = dx^\mu dx_\mu + dz_3^2 + H (dy^2 + dz_2^2 +
dz_4^2) + H^{-1} (dz_1 + {\cal A}_1^{(1)})^2 \ ,
\nonumber\\
{\cal A}_0^{(13)}:&& ds_{11}^2 = dx^\mu dx_\mu + dz_2^2 + H (dy^2 + dz_3^2 +
dz_4^2) + H^{-1} (dz_1 + {\cal A}_1^{(1)})^2 \ ,
\nonumber\\
{\cal A}_0^{(23)}:&& ds_{11}^2 = dx^\mu dx_\mu + dz_1^2 + H (dy^2 + dz_3^2 +
dz_4^2) + H^{-1} (dz_2 + {\cal A}_1^{(2)})^2 \ ,
\label{7ddox}\\
A_0^{(123)}:&& ds_{11}^2 = H^{-\ft13} dx^\mu dx_\mu + H^{\ft23} 
(dy^2 + dz_1^2 + dz_2^2 +dz_3^2 +dz_4^2) \ ,\nonumber
\eea
where we indicate which axion is used in the \ss reduction for each case. In
the first three metrics given here, the gauge potential is given by ${\cal
A}_1^{(1)}= m z_4\, dz_2$, ${\cal A}_1^{(1)}= m z_4\, dz_3$ and ${\cal
A}_1^{(2)}= m z_4\, dz_3$ respectively.  The last metric, corresponding to
the oxidation of the domain wall solution of the $A_0^{(123)}$  massive
theory in $D=7$, has a straightforward eleven-dimensional interpretation as
a continuous 4-volume of 5-branes. The previous cases, which all come from
solutions where the axion used in the \ss reduction originates from the
metric, give rise to ``twisted'' solutions in $D=11$. 

     The 5-brane solution of the sixth massive $D=7$ supergravity theory,
which comes from the \kk dimensional reduction of the massive IIA theory in
$D=10$, can be oxidised back to the 8-brane \cite{pw} of the massive IIA
theory, or to a stack of 7-branes in the type IIB theory in $D=10$
\cite{bdgpt}. 

     As we saw in section 3, it is also possible to construct 5-branes with
$\Delta=2$ in $D=7$.  These will oxidise back to appropriate intersections
of the $\Delta=4$ oxidations described above. 

     As one descends through the dimensions, similar considerations can be
applied to all of the domain-wall solutions to the multitude of massive
supergravities.  One general feature that is worth remarking upon is that
the only values of $\Delta$ that can arise in any of the massive
supergravities obtainable from $D=11$ are the set $\Delta=4/N$, as discussed
above in section 3.  In particular, this means that if a massive
supergravity theory has a single-scalar truncation with any value of
$\Delta$ other than one of these allowed ones, then one can immediately
deduce that it cannot come from $D=11$ by \ss dimensional reduction. 
Examples of such independent massive supergravities are the gauged
supergravities in $D=7$ \cite{tn}, $D=6$ \cite{at,rom} and $D=5$ \cite{gst},
and the Freedman-Schwarz gauged theory in $D=4$ \cite{fs}. Interestingly,
these all have $\Delta=-2$.  While it is conceivable therefore that they
might all have a common origin in the gauged $D=7$ theory, it must certainly
be the case that none of them can come from $D=11$ supergravity by \ss
reduction. 

\section{Discussion and conclusions}

     In this paper, we have studied the massive supergravity theories that
can be obtained by \ss dimensional reduction from $D=11$ supergravity. This
is a generalisation of the usual \kk dimensional reduction, in which at one
stage of the step-by-step reduction from eleven to $D$ dimensions, one or
more of the axions in the theory in an intermediate dimension are allowed to
have a linear dependence on the compactification coordinate $z$, of the form
given in (\ref{ssans}).  This generalised ansatz still yields a consistent
truncation of the theory, because the axions in question are covered by
derivatives everywhere in the Lagrangian, and thus the Lagrangian will be
independent of $z$ after substitution of the ansatz. The ansatz has the
effect of generating cosmological-type terms after the dimensional
reduction, together with mass terms for certain fields in the theory. 

     We have seen that there are many different massive supergravities that
can be obtained by this procedure.  This contrasts with the situation for
the usual \kk reduction of $D=11$ supergravity, where one obtains just one
massless maximal supergravity in each dimension.  The reason for this
multiplicity of massive theories is that the \ss reduction step commutes
neither with ordinary \kk reduction, nor with U-duality.  Thus, for example,
we obtain a single massive supergravity in $D=8$ by this method, since there
is only one axion in $D=9$ supergravity, but in $D=7$ we obtain in total
five different massive supergravities.  Four of these correspond to the four
different axions in $D=8$ that can be used for the \ss reduction, and the
fifth is the theory obtained by ordinary \kk reduction of the massive theory
we already constructed in $D=8$.  In addition, there is always one further
massive maximal supergravity in each dimension $D\le 10$, corresponding to
the massive IIA supergravity \cite{r} and its \kk dimensional reductions. 
This latter sequence of theories is quite distinct from any of those
obtainable from $D=11$ supergravity by the methods used in this paper.  In 
particular, the conjectured duality \cite{bdgpt} in $D=8$ of the
vertically-reduced 6-brane of massless type IIA supergravity and the
diagonally-reduced 8-brane of massive type IIA supergravity seems to be
false. It is intriguing, though, that the dimensional reduction of the
massive IIA theory to $D=9$ is the same as the theory one obtains by the \ss
reduction of the IIB theory in $D=10$ \cite{bdgpt}. 

    The distinguished r\^ole that the massive type IIA theory appears to
play, amongst the set of all massive theories, suggests the possibility that 
it may have its origins in some more fundamental theory.  This would be in a 
similar spirit to the introduction of M-theory, which is dual to the 
massless type IIA string, and F-theory, which is dual to the type IIB 
string.  In this second example, the $SL(2,Z)$ symmetry of the type IIB 
theory can then be understood as coming from the geometry of the 2-torus 
used to compactify from $D=12$ to $D=10$ \cite{v}.  Thus we might
conjecture the existence of an hypothetical H-theory, whose compactification
would give rise to the massive IIA theory in $D=10$.  As we have seen, a
massive supergravity can be obtained by performing an $S^1$ reduction of a
theory in one higher dimension, by allowing an axion, or 0-form potential,
to depend linearly on the compactification coordinate.  The most natural
source of such an axion in $D=11$ would be from the dimensional reduction of
the metric in a theory in $D=13$. (The existence of a theory in $D=13$ has
also been proposed in \cite{b}.)  Thus we may imagine that there are
three fundamental theories, namely M, F and H in $D=11$, 12 and 13, which
are the progenitors of the IIA, IIB and massive IIA theories respectively. 
Just as the IIA and IIB theories compactified on $S^1$ are T-duals, so the
IIB and massive IIA compactified on $S^1$ are T-duals (where in this latter 
case, the $S^1$ compactification of the IIB theory involves a linear 
dependence of the axion on the compactification coordinate).

     There is in fact another way to interpret the \ss dimensional reduction
that we have been carrying out in this paper.  Let us suppose that we have a
formulation of the massless theory in $D+1$ dimensions in which the axion
$\chi$ is covered everywhere by a derivative.  Since it therefore appears in
the Lagrangian only {\it via} its 1-form field $F_{\sst D}=\,{}^*\!F_1$, we
may rewrite the Lagrangian in a dualised form in terms of the $D$-form field
strength $F_{\sst D}=\,{}^*\!F_1$.  Upon now performing a standard \kk
reduction, in which the potential for $F_{\sst D}$ is re-expressed in terms
of $z$-independent $D$-dimensional potentials $A_{\sst D-1}$ and $A_{\sst
D-2}$ as in (\ref{kkans}), the $D$-dimensional Lagrangian will acquire
kinetic terms for field strengths $F_{\sst D}$ and $F_{\sst D-1}$ of ranks
$D$ and $(D-1)$.  If these are then dualised in $D$ dimensions, the first
will give rise to a cosmological-type term, and the second will give a
1-form field strength that can be interpreted as the derivative of an axion.
This is the same result that we obtained directly by simply performing the
\ss reduction of the axion in $(D+1)$ dimensions.  Clearly, this alternative
description can also be generalised to the case where more than one axion is
covered by derivatives simultaneously, and thus we have a general statement
that for each such axion there is a correspondence between a \ss reduction
step on the one hand, and a dualisation, followed by a \kk reduction step,
followed by a dualisation in the lower dimension. From our point of view,
however, it should be emphasised that this alternative description is not
completely equivalent:  after dualising the 1-form field strength of a
0-form axion, and performing a \kk reduction, one cannot give a local
expression for the higher-dimensional 0-form in terms of the
lower-dimensional fields.  Since we are interested in formulating the
dimensional reduction process in such a way that all solutions of the
lower-dimensional theories are re-interpreted as solutions of $D=11$
supergravity, it is important that one should be able to give the \kk ansatz
for the $D=11$ metric and 3-form potential in terms of the fields of the
lower-dimensional theories, and thus dualisations at intermediate stages of
the reduction should be avoided. 

     For the same reason, the potential $A_3$ in $D=5$, and the potentials
$A_2^{(i)}$ in $D=4$,  should be kept in these original forms, and should
not be reformulated as axions by performing dualisations.  Thus these fields
will undergo standard \kk reductions to $D=4$ and $D=3$ respectively, giving
rise to (non-propagating) $D$-form descriptions of massive supergravities,
rather than the dual descriptions with cosmological terms. It is interesting
to note that, historically, the first example of a massive supergravity
theory was of just such a kind. Specifically, a massive theory in $D=4$ was
obtained by performing the \kk dimensional reduction of the 4-form field
strength $F_4$ in $D=5$, and then dualising the resulting 4-form in $D=4$ to
give a cosmological-type term \cite{ant}. 

    It is perhaps worth remarking, in the light of the above discussion,
that one can in fact reformulate the massive IIA theory in ten dimensions in
terms of a 10-form field strength rather than a cosmological-type term
\cite{bdgpt}.  Effectively, the mass parameter of the massive IIA theory now
arises as a constant of integration.  In particular, this constant of
integration can take the value zero, implying that all solutions of the
massless IIA theory are also solutions of the 10-form reformulation of the
massive IIA theory.  Since all of our massive theories that are obtainable
from D=11 pass through $D=10$ without yet having performed any \ss
reduction, it follows that they can equally well be viewed as descendants of
the massless IIA theory in $D=10$. Then, by virtue of the previous
observation, we may say that they are in fact descendants of the 10-form
reformulation of the massive IIA theory in $D=10$.  In this sense, one may
in fact take the point of view that this formulation of the massive IIA
theory provides a ``universal'' theory from which {\it all} the
lower-dimensional massive theories are derivable.  They can of course also
therefore be ``derived'' from the original formulation of the massive IIA
theory with a cosmological term, in that those massive theories that we have
obtained from $D=11$ can also be obtained from the massive IIA theory by
first setting its mass parameter to zero.  However, although this gives an
algorithm for generating theories, their solutions would not be solutions of
the massive IIA theory in its original form, since its mass parameter is a
given, fixed quantity, and not an arbitrary integration constant that can
take any value including zero. 

     The above observations do not alter the fact that there exist
inequivalent maximally-supersymmetric massive theories in each dimension
$D\le 8$.  In particular, the theory that one gets by performing a \ss
reduction on any one axion is inequivalent to the theory one gets by
performing the \ss reduction on any other axion.   There are, as we have
seen, larger theories with multiple parameters that correspond to performing
simultaneous \ss reductions on more than one axion.  However, it is not
possible to perform \ss reductions on all the axions simultaneously, and
thus there does not in general exist any single universal theory in a lower
dimension that encompasses all of the single-axion reductions at the same
time. 

\renewcommand{\theequation}{A.\arabic{equation}}
\section*{Appendix A.}

The dimensional reduction of the bosonic sector of $D=11$ supergravity to
$D$ dimensions, in the formalism that we are using in this paper, is given by
\cite{lpsol}
\bea
e^{-1}{\cal L} &=& R -\ft12 (\del\vec\phi)^2 -\ft1{48} e^{\vec a\cdot \vec
\phi}\, F_4^2 -\ft{1}{12} \sum_i e^{\vec a_i\cdot \vec\phi}\, (F_3^{i})^2
-\ft14 \, \sum_{i<j} e^{\vec a_{ij}\cdot \vec\phi}\, (F_2^{ij})^2
\label{dgenlag}\\
&& -\ft14 \sum_i e^{\vec b_i\cdot \vec\phi}\, ({\cal F}_2^i)^2
-\ft12  \sum_{i<j<k} e^{\vec a_{ijk} \cdot\vec \phi}\,
(F_1^{ijk})^2 -\ft12 \sum_{i<j} e^{\vec b_{ij}\cdot \vec\phi}\,
({\cal F}_1^{ij})^2 + e^{-1} {\cal L}_{\sst{FFA}}\ ,\nonumber
\eea
where $F_4$, $F_3^i$, $F_2^{ij}$ and $F_1^{ijk}$ are the 4-form, 3-forms,
2-forms and 1-forms coming from the dimensional reduction of $\hat F_4$ in
$D=11$; ${\cal F}_2^i$ are the 2-forms coming from the dimensional reduction
of the vielbein, and ${\cal F}_1^{ij}$ are the 1-forms coming from the
dimensional reduction of these 2-forms.  The axions that are the focus of
our attention in much of this paper are the 0-form potentials $A_0^{ijk}$
and ${\cal A}_0^{ij}$. 

     The $(11-D)$ scalar fields denoted by $\vec\phi$ are the dilatonic
scalars arising at each step in the sequential \kk reduction of the metric. 
The $\vec a$ and $\vec b$ vectors appearing in the exponentials are
constants characterising the couplings of the dilatonic scalars to the
various field strengths.  They are given by \cite{lpsol} 
\bea
&&F_{\sst{MNPQ}}\qquad\qquad\qquad\qquad\qquad\qquad\qquad\qquad
{\rm vielbein}\nonumber\\
{\rm 4-form:}&&\vec a = -\vec g\ ,\nonumber\\
{\rm 3-forms:}&&\vec a_i = \vec f_i -\vec g \ ,\nonumber\\
{\rm 2-forms:}&& \vec a_{ij} = \vec f_i + \vec f_j - \vec g\ ,
\qquad\qquad\qquad\qquad\qquad \,\,\, \,\vec b_i = -\vec f_i \ ,
\label{dilatonvec}\\
{\rm 1-forms:}&&\vec a_{ijk} = \vec f_i + \vec f_j + \vec f_k -\vec g
\ ,\qquad\qquad\qquad\qquad\vec b_{ij} = -\vec f_i + \vec f_j\ ,\nonumber \\
{\rm 0-forms:}&& \vec a_{ijk\ell} =\vec f_i +\vec f_j+\vec f_k +\vec f_\ell
-\vec g \ ,\qquad\qquad\quad\ \  \vec b_{ijk}=-\vec f_i +\vec f_j +\vec f_k\ ,
\nonumber
\eea
where the vectors $\vec g$ and $\vec f_i$ have $(11-D)$ components
in $D$ dimensions, and are given by
\bea
\vec g &=&3 (s_1, s_2, \ldots, s_{11-\sst D})\ ,\nonumber\\
\vec f_i &=& \Big(\underbrace{0,0,\ldots, 0}_{i-1}, (10-i) s_i, s_{i+1},
s_{i+2}, \ldots, s_{11-\sst D}\Big)\ ,\label{gfdef}
\eea
where $s_i = \sqrt{2/((10-i)(9-i))}$.  It is easy to see that they satisfy
\be
\vec g \cdot \vec g = \ft{2(11-D)}{D-2}, \qquad
\vec g \cdot \vec f_i = \ft{6}{D-2}\ ,\qquad
\vec f_i \cdot \vec f_j = 2\delta_{ij} + \ft2{D-2}\ .\label{gfdot}
\ee
We have included the dilaton vectors for ``0-form field strengths'' in
(\ref{dilatonvec}) because these fit into the same general pattern,
and because they correspond to the cosmological-type terms arising in
the \ss reductions discussed in this paper.

     In general, the field strengths appearing in the kinetic terms acquire
Chern-Simons type modifications in the dimensional reduction process, given
by \cite{lpsol} 
\bea
F_4 &=& \td F_4 - \gamma^{ij} \td F_3^i\wedge {\cal A}_1^j -\ft12
\gamma^{ik}\gamma^{j\ell} \td F_2^{ij} \wedge {\cal A}_1^k\wedge
{\cal A}_1^\ell + \ft16 \gamma^{i\ell}\gamma^{jm}\gamma^{kn}
\td F_1^{ijk}\wedge {\cal A}_1^\ell \wedge {\cal A}_1^m \wedge
{\cal A}_1^n\ ,\nonumber\\
F_3^i &=& \gamma^{ji}\td F_3^j - \gamma^{ji}\gamma^{k\ell} \td F_2^{jk}
\wedge {\cal A}_1^\ell - \ft12 \gamma^{ji}\gamma^{km}\gamma^{\ell n}
\td F_1^{jk\ell}\wedge {\cal A}_1^m \wedge {\cal A}_1^n\ ,\nonumber\\
F_2^{ij} &=& \gamma^{ki}\gamma^{\ell j} \td F_2^{k\ell} -
\gamma^{ki} \gamma^{\ell j} \gamma^{mn} \td F_1^{k\ell m}\wedge
{\cal A}_1^n\ ,\nonumber\\
F_1^{ijk} &=& \gamma^{\ell i} \gamma^{mj} \gamma^{nk} \td F_1^{\ell mn}
\ ,\label{A.6}\\
{\cal F}_2^i &=& \td {\cal F}_2^i - \gamma^{jk} \td {\cal F}_1^{ij} \wedge
{\cal A}_1^k\ ,\nonumber\\
{\cal F}_1^{ij} &=& \gamma^{kj} \td {\cal F}_1^{ik}\ ,
\eea
where the tilded quantities denote the unmodified field strengths given
directly by the exterior derivatives of gauge potentials.  The matrix
$\gamma^{ij}$ is given by
\be
\gamma^{ij}=\Big[(1+{\cal A}_0)^{-1}\Big]^{ij}=
\delta^{ij} - {\cal A}_0^{ij} + {\cal A}_0^{ik} {\cal A}_0^{kj}
-{\cal A}_0^{ik} {\cal A}_0^{k\ell} {\cal A}_0^{\ell j} + \cdots\ .\label{gam}
\ee
The 0-form potentials ${\cal A}_0^{ij}$ are defined only for $i<j$,
{\it i.e.}\ ${\cal A}_0^{ij}=0$ for $i\ge j$, and hence it follows that 
the series (\ref{gam}) for $\gamma^{ij}$ terminates after 
$j-i+1\le 11-D$ terms, with 
$\gamma^{ij}=0$ for $i>j$ and $\gamma^{ij} = 1$ for $i=j$. 

     The term ${\cal L}_{\sst{FFA}}$ in (\ref{dgenlag}) is the dimensional 
reduction of the $\td F_4\wedge\td F_4\wedge A_3$ term in $D=11$, and is
given in lower dimensions by \cite{lpsol}
\bea
D=10: &&\ft12 \td F_4\wedge \td F_4 \wedge A_2\ ,\nonumber\\
D=9: &&\Big(-\ft14 \td F_4 \wedge \td F_4 \wedge A_1^{ij}-\ft12 \td F_3^i
\wedge \td F_3^j \wedge A_3\Big)\epsilon_{ij}\ ,\nonumber\\
D=8: && \Big(-\ft1{12} \td F_4\wedge \td F_4 A_0^{ijk} -\ft16 \td F_3^i\wedge
\td F_3^j \wedge A_2^k +\ft12 \td F_3^i \wedge \td F_2^{jk} \wedge
A_3\Big) \epsilon_{ijk}\ ,\nonumber\\
D=7: && \Big(-\ft16 \td F_4\wedge \td F_3^i A_0^{jkl} +\ft16 \td F_3^{i}\wedge
\td F_3^{j} \wedge A_1^{kl} +\ft18 \td F_2^{ij}\wedge \td F_2^{kl}
\wedge A_3\Big)\epsilon_{ijkl}\ ,\label{ffaterms}\\
D=6: && \Big(\ft1{12} \td F_4\wedge \td F_2^{ij} A_0^{klm} +\ft1{12}
\td F_3^i\wedge \td F_3^j A_0^{klm} +\ft18 \td F_2^{ij}\wedge
\td F_2^{kl} \wedge A_2^m\Big) \epsilon_{ijklm}\ ,\nonumber\\
D=5: && \Big(\ft1{12} \td F_3^i\wedge \td F_2^{jk}  A_0^{lmn} -\ft1{48}
\td F_2^{ij}  \wedge \td F_2^{kl}\wedge A_1^{mn} -\ft1{72}
\td F_1^{ijk}\wedge \td F_1^{lmn} \wedge A_3\Big)
\epsilon_{ijklmn}\ ,\nonumber\\
D=4: && \Big(-\ft1{48} \td F_2^{ij}\wedge \td F_2^{kl} A_0^{mnp} -\ft1{72}
\td F_1^{ijk}\wedge \td F_1^{lmn} \wedge A_2^p\Big)
\epsilon_{ijklmnp}\ ,\nonumber\\
D=3:&& \ft1{144}\, \td F_1^{ijk}\wedge \td F_1^{lmn}\wedge A_1^{pq}
\epsilon_{ijklmnpq}\ ,\nonumber\\
D=2: && \ft1{1296}\, \td F_1^{ijk}\wedge \td F_1^{lmn} A_0^{pqr}
\epsilon_{ijklmnpqr}\ .\nonumber
\eea

\section*{Acknowledgement}

       K.S.S. would like to thank S.I.S.S.A. and the Yukawa Institute,
University of Kyoto, for hospitality during the course of this work.

\end{document}